\documentclass[11pt,twoside,nohyper]{pallis}
\usepackage{epsfig}
\usepackage{latexsym}
\usepackage{euscript}
\usepackage{amssymb}
\usepackage{pstricks}
\usepackage{fancyhed}

\newcommand{\beq}{\begin{equation}}
\newcommand{\eeq}{\end{equation}}
\newcommand{\ben}{\begin{equation*}}
\newcommand{\een}{\end{equation*}}
\newcommand{\ba}{\begin{eqnarray}}
\newcommand{\ea}{\end{eqnarray}}
\newcommand{\ban}{\begin{eqnarray*}}
\newcommand{\ean}{\end{eqnarray*}}
\newcommand{\brr}{\begin{array}}
\newcommand{\err}{\end{array}}
\newcommand{\bc}{\begin{center}}
\newcommand{\ec}{\end{center}}

\newcommand{\bea}{\begin{eqnarray}}
\newcommand{\eea}{\end{eqnarray}}
\newcommand{\bean}{\begin{eqnarray*}}
\newcommand{\eean}{\end{eqnarray*}}
\newcommand{\ftn}{\footnotesize}
\newcommand{\nsz}{\normalsize}
\newcommand{\ssz}{\scriptsize}

\title{\huge M{\Large ASSIVE} P\kern-3.5pt{\Large ARTICLE} D{\Large ECAY AND \\}
C{\Large OLD} D\kern-1.pt{\Large ARK} M\kern-1.pt{\Large ATTER}
A\Large BUNDANCE}

\author{\Large C. P\kern-1.5pt\nsz ALLIS\\ SISSA/ISAS, \\ Via Beirut
2-4,\\ 34013 Trieste, ITALY\\ \vspace{0.05in}
Physics Division, School of Technology,\\
Aristotle University of Thessaloniki, \\
541 24 Thessaloniki, GREECE \\ \vspace{5pt}
\email{kpallis@auth.gr}}

\abstract{\\ $~~~$ The decoupling of a cold relic, during a
decaying-particle-dominated cosmological evolution is analyzed,
the relic density is calculated both numerically and
semi-analytically and the results are compared with each other.
Using plausible values (from the viewpoint of supersymmetric
models) for the mass and the thermal averaged cross section times
the velocity of the cold relic, we investigate scenaria of
equilibrium or non-equilibrium production. In both cases,
acceptable results for the dark matter abundance can be obtained,
by constraining the reheat temperature of the decaying particle,
its mass and the averaged number of the produced cold relics. The
required reheat temperature is, in any case, lower than about
$20~{\rm GeV}$.

\\ \\{\sc Keywords}: Cosmology, Dark Matter \\ {\sc PACS codes}: 98.80.Cq, 95.35.+d
\\ \\ {\sl\bfseries Published in} {\sl Astropart. Phys.} {\bf 21}, 689 (2004)}

\begin{document}

\setcounter{page}{1} \pagestyle{fancyplain}

\addtolength{\headheight}{.5cm}

\rhead[\fancyplain{}{ \bf \thepage}]{\fancyplain{}{M{\ftn ASSIVE}
P{\ftn ARTICLE} D{\ftn ECAY AND} C{\ftn OLD} D{\ftn ARK} M{\ftn
ATTER} A{\ftn BUNDANCE}}} \lhead[\fancyplain{}{
\leftmark}]{\fancyplain{}{\bf \thepage}} \cfoot{}

\section{I{\ftn NTRODUCTION}}\label{sec:intro}

\hspace{.562cm} The enigma of the Cold Dark Matter (CDM)
constitution of the universe becomes more and more precisely
defined, after the recently announced WMAP results \cite{wmap,
wmapl}, which determine the CDM abundance, $\Omega_{\rm CDM}h^2$,
with an unprecedented accuracy:
\beq \Omega_{\rm CDM}h^2=0.1126_{-0.0181}^{+0.0161}
\label{cdmba}\eeq
at $95\%$ confidence level. In light of this, the relic density of
any CDM candidate $\tilde\chi$ (i.e, which decouples being non
relativistic), $\Omega_{\tilde\chi}h^2$, is to satisfy a very
narrow range of values:
\beq {\sf (a)}~~0.09\lesssim
\Omega_{\tilde\chi}h^2~~\quad\mbox{and}\quad~~ {\sf
(b)}~~\Omega_{\tilde\chi}h^2\lesssim0.13 \label{cdmb}\eeq
which tightly restricts the parameter space of the theories which
support the existence of $\tilde\chi$. The most popular of these
are the $R$-parity conserving supersymmetric ({\sc SUSY}) theories
which identify $\tilde\chi$ to the stable lightest {\sc SUSY}
particle ({\sc LSP}) \cite{goldberg}. According to the standard
scenario, (i) $\tilde\chi$ decouples from the cosmic fluid during
the radiation-dominated (RD) era, (ii) being in chemical
equilibrium with plasma and (iii) produced through thermal
scatterings in the plasma. The condition (ii) is satisfied
naturally by the lightest neutralino of the minimal {\sc SUSY}
standard model ({\sc MSSM}), which turns out to have the required
strength of interactions, being weakly interacting.

Although quite compelling, this scenario comes across with
difficulties, especially when it is applied in the context of
economical and predictive versions of {\sc MSSM}. E.g., in most of
the parameter space of the Constrained {\sc MSSM} ({\sc CMSSM})
\cite{Cmssm}, $\tilde\chi$ turns out to be bino and its
$\Omega_{\tilde\chi}h^2$ exceeds the bound of Eq. (\ref{cdmb}{\sf
b}). Several suppression mechanisms of $\Omega_{\tilde\chi}h^2$
have been proposed so far: Bino-sleptons \cite{ellis2} and
particularly, for large $\tan\beta$, bino-stau \cite{cdm},
bino-stops \cite{boem, santoso1} or bino-chargino \cite{darkn}
coannihilations and/or A-pole effect \cite{lah} can efficiently
reduce $\Omega_{\tilde\chi}h^2$. Also several kinds of
non-universality in the Higgs \cite{ellis3} and/or gaugino
\cite{edjo, nelson} and/or sfermionic sector \cite{su5b} can help
in the same direction, creating additional coannihilation effects.
As is expected, a more or less tuning of the {\sc SUSY} parameters
is needed in these cases, without a simultaneous satisfaction of
other phenomenological constraints to be always possible (see,
e.g. Ref. \cite{wmapl}). On the other hand,
$\Omega_{\tilde\chi}h^2$ turns out to be lower than the bound of
Eq. (\ref{cdmb}{\sf a}) in other models, e.g., in the anomaly
mediated {\sc SUSY} breaking model ({\sc AMSBM}) where
$\tilde\chi$ is mostly wino \cite{wells} or in models based on
$SU(5)$ gaugino non-universality \cite{roy}, where $\tilde\chi$
can be higgsino. Then we have to invoke another CDM candidate
\cite{axino}, in order for the range of Eq. (\ref{cdmb}) to be
fulfilled.

However, this picture can dramatically change, if the standard
assumption (i) is lifted. Indeed, since there is no direct
information for the history of the Cosmos before the epoch of
nucleosynthesis (i.e., temperatures $T>1~{\rm MeV}$) the
decoupling of $\tilde\chi$ can occur not in the RD era. As was
pointed out a lot years ago \cite{McDonald} and, also, recently
\cite{riotto, fornengo}, the ${\tilde\chi}$ decoupling can be
related to the decay of a massive scalar particle. The modern
cosmo-particle theories are abundant in such fields, e.g.
inflatons \cite{chung, kubo, drees, dreesa}, dilatons or moduli
\cite{quevedo, moroi}, Polonyi field \cite{yanagida}, $q$-balls
\cite{fujii} (see, also \cite{rachel}). During their decay, these
particles perform coherent oscillations ``reheating'' the universe
\cite{turner}. This phenomenon is not instantaneous \cite{kolb,
chung}. In particular, the maximum temperature during this period
is much larger than the so-called reheat temperature, which can be
better considered as the largest temperature of the RD era
\cite{riotto}. Consequently, the ``freeze out'' of $\tilde\chi$
could be realized before the completion of the reheating. The
cosmological evolution during this phase is strongly modified as
regards the standard one \cite{turner}, with crucial consequences
to $\Omega_{\tilde\chi}h^2$ calculation \cite{riotto, fornengo,
moroi, shaaban}. Namely, two types of $\tilde\chi$-production
emerge, in contrast with (ii): The chemically equilibrium (EP) and
the non-equilibrium production (non-EP) (in both cases, kinetic
equilibrium of $\tilde\chi$'s is assumed \cite{riotto}). In this
paper we extend the analysis in Ref. \cite{riotto}, lifting also
the assumption (iii) of the standard scenario: We include the
possibility (which, naturally arises even without direct coupling
\cite{drees, dreesa}) that the decaying particle can decay to
$\tilde\chi$. The problem has already been faced
semi-quantitatively in Refs \cite{moroi, shaaban, drees, dreesa}
and numerically in Ref. \cite{brazil} for significantly more
massive $\tilde\chi$'s.

Our numerical and semi-analytical analyses are exposed in secs
\ref{sec:num} and \ref{sec:neut}. The obtained results are
compared with each other in sec. \ref{appl}. There, we realize,
also, a model independent application of our findings in the case
of {\sc SUSY} models inspired ${\tilde\chi}$ masses and cross
sections. We find that comfortable satisfaction of Eq.
(\ref{cdmb}) can be achieved, by constraining the reheat
temperature to rather low values, the mass of the decaying
particle and the averaged number of the produced $\tilde\chi$'s,
without any tuning of the {\sc SUSY} parameters.

Throughout the text and the formulas, brackets are used by
applying disjunctive correspondence and natural units
($\hbar=c=k_B=1$) are assumed.

\section{D{\ftn YNAMICS OF}  M{\ftn ASSIVE} P{\ftn ARTICLE}
D{\ftn ECAY}} \label{sec:num}

\hspace{.562cm} We consider a scalar particle $\phi$ with mass
$m_\phi$, which decays with a rate $\Gamma_\phi$ into radiation,
producing an average number $N_{\tilde\chi}$ of $\tilde\chi$'s
with mass $m_{\tilde\chi}$, rapidly thermalized. We, also, let
open the possibility (contrary to Ref. \cite{moroi}) that
${\tilde\chi}$'s are produced through thermal scatterings in the
bath. Our theoretical analysis is presented in sec. \ref{Beqs} and
its numerical treatment in sec. \ref{Neqs}. Useful approximated
expressions are derived in sec. \ref{Seqs}.

\subsection{R{\ssz ELEVANT} B{\ssz OLTZMANN} E{\ssz QUATIONS}}
\label{Beqs}

\hspace{.562cm} The energy density of radiation $\rho_{_{\rm R}}$
and the number densities of $\phi$, $n_\phi$, and $\tilde\chi$,
$n_{\tilde\chi}$, satisfy the following Boltzmann equations
\cite{chung, moroi} (we use the shorthand
$\Delta_\phi=(m_\phi-N_{\tilde\chi}m_{\tilde\chi})/m_\phi$):
\begin{eqnarray}
&& \dot n_\phi+3Hn_\phi+\Gamma_\phi n_\phi=0,\label{nf} \\
&& \dot \rho_{_{\rm R}}+4H\rho_{_{\rm R}}-\Gamma_\phi \Delta_\phi
m_\phi n_\phi-2m_{\tilde\chi}\langle \sigma v \rangle \left(
n_{\tilde\chi}^2 - n_{\tilde\chi}^{\rm eq2}\right)=0, \label{rR}\\
&& \dot n_{\tilde\chi}+3Hn_{\tilde\chi}+\langle \sigma v \rangle
\left( n_{\tilde\chi}^2 - n_{\tilde\chi}^{\rm
eq2}\right)-\Gamma_\phi N_{\tilde\chi} n_\phi=0,\label{nx}
\end{eqnarray}
where dot stands for derivative with respect to (w.r.t) the cosmic
time $t$, $\langle \sigma v \rangle$ is the thermal-averaged cross
section of $\tilde\chi$ particles times velocity and $H$ the
Hubble expansion parameter which is given by ($M_{\rm
P}=1.22\times10^{19}~{\rm GeV}$ is the Planck scale):
\begin{equation} \label{Hini}
H^2=\frac{8\pi}{3M_{\rm P}^2} \left(\rho_\phi +\rho_{_{\rm R}}+
\rho_{\tilde\chi}
\right),\quad\mbox{with}\quad\rho_\phi=\Delta_\phi m_\phi
n_\phi\quad\mbox{and}\quad \rho_{\tilde\chi}= m_{\tilde\chi}
n_{\tilde\chi}.
\end{equation}
The equilibrium number density of $\tilde\chi$,
$n_{\tilde\chi}^{\rm eq}$ obeys the Maxwell-Boltzmann statistics:
\begin{equation} \label{neq}
n_{\tilde\chi}^{\rm eq}(x)=\frac{g}{(2\pi)^{3/2}}
m_{\tilde\chi}^3\>x^{3/2}\>e^{-1/x}P_2(1/x),\quad\mbox{where}\quad
x=T/m_{\tilde\chi},
\end{equation}
$g=2$ is the number of degrees of freedom of ${\tilde\chi}$ and
$P_n(z)=1+(4n^2-1)/8z$ is obtained by asymptotically expanding the
modified Bessel function of the second kind of order $n$.

The temperature, $T$, and the entropy density, $s$, can be found
using the relations:
\beq \mbox{\sf (a)}~~\rho_{_{\rm R}}=\frac{\pi^2}{30}g_{\rho*}\
T^4\quad \mbox{and}\quad \mbox{\sf
(b)}~~s=\frac{2\pi^2}{45}g_{s*}\ T^3, \label{rs}\eeq
where $g_{\rho*}(T)~[g_{s*}(T)]$ are the energy [entropy]
effective number of degrees of freedom at temperature $T$. Their
numerical values are evaluated by using the tables included in
{\tt micrOMEGAs} \cite{micro}, originated from the {\sf DarkSUSY}
package \cite{dark}.

Central role to our investigation plays the reheat temperature.
Its precise value, $T_{\rm rh}$,  can be found, by solving
numerically the following \cite{lazarides}:
\beq \rho_{_{\rm R}}(T_{\rm rh})=\rho_\phi(T_{\rm rh}).
\label{Trhp}\eeq
However, following Ref. \cite{riotto}, we prefer to handle reheat
temperature as an input parameter, $T_{\rm RH}$, defining it
through an analytic formula, which, however approximates fairly
(within 10$\%$) $T_{\rm rh}$. This can be expressed in terms of
$\Gamma_\phi$, using the following \cite{kolb, lazarides}:
\begin{equation}
\Gamma_\phi =4\sqrt{\frac{\pi^3 g_{\rho*}(T_{\rm RH})}{45}}
\frac{T_{\rm RH}^2}{M_{\rm P}} \label{GTrh} \cdot
\end{equation}
Eq. (\ref{GTrh}) is derived consistently with Eq. (\ref{Hear}) and
differs from the corresponding definition in Ref. \cite{riotto} by
a factor 2 (our choice is justified in sec. \ref{thnonth}).

\subsection{N{\ssz UMERICAL} I{\ssz NTEGRATION}}
\label{Neqs}

\hspace{.562cm} The numerical integration of Eqs
(\ref{nf})--(\ref{nx}) is facilitated by absorbing the dilution
terms. To this end, we convert the time derivatives to derivatives
w.r.t the scale factor, $R$ \cite{chung, riotto}. We find it
convenient to define the following dimensionless variables:
\begin{equation} \label{fdef}
f_\phi=n_\phi R^3, \quad f_{\rm R}=\rho_{_{\rm
R}}R^4\quad\mbox{and}\quad f_{\tilde\chi}^{[\rm eq]}=n^{[\rm
eq]}_{\tilde\chi} R^3\ .
\end{equation}
In terms of these variables, Eqs\ (\ref{nf})--(\ref{nx}) become:
\begin{eqnarray}
\frac{df_\phi}{dR} &=&-\Gamma_\phi\frac{f_\phi}{H R},\label{ff}\\
\frac{df_{\rm R}}{dR} &=&\Gamma_\phi\Delta_\phi
m_\phi\frac{f_\phi}{H}+2 m_{\tilde\chi} \langle \sigma v \rangle
\frac{ f_{\tilde\chi}^2 - f_{\tilde\chi}^{\rm eq2}}{H R^3},
\label{fR}
\\
\frac{df_{\tilde\chi}}{dR} &=& -\langle\sigma v\rangle
\frac{f_{\tilde\chi}^2 - f_{\tilde\chi}^{\rm eq2}}{H R^4
}+\Gamma_\phi N_{\tilde\chi}\frac{f_\phi}{HR},\label{fn}
\end{eqnarray}
where
\begin{equation} \label{H2exp}
H=\sqrt{\frac{8\pi}{3}}\frac{R^{-3/2}}{M_{\rm P}}
\left(\Delta_\phi m_\phi f_\phi +f_{\rm R}/R + m_{\tilde\chi}
f_{\tilde\chi} \right)^{1/2} \ .\end{equation}
Since at early times the energy density of the universe is
completely dominated by this of $\phi$ field, $m_\phi n_{_{\rm
I}}$, the system of Eqs\ (\ref{ff})--(\ref{fn}) is solved,
imposing the following initial conditions:
\begin{equation}
f_\phi(R_{\rm I}) =n_{_{\rm I}} R^3_{\rm I}\quad\mbox{and} \quad
f_{\rm R}(R_{\rm I})=f_{\tilde\chi}(R_{\rm I})=0,\quad
\mbox{with}\quad R_{\rm I}=m_{\phi}^{-1}\ . \label{init}
\end{equation}
\subsection{S{\ssz EMI}-A{\ssz NALYTICAL} A{\ssz PPROACH}}
\label{Seqs}

\hspace{.562cm} We can obtain a comprehensive and rather accurate
approach of the dynamics of the decaying-particle-dominated
cosmology, following the arguments of Ref. \cite{riotto}. At the
epoch before the completion of reheating, $T\gg T_{\rm RH}$, the
universe is dominated by the number density of $\phi$ which
initially has a large value $n_{_{\rm I}}$. Its evolution is given
by the exact solution of Eq. (\ref{nf}) \cite{turner, kolb,
lazarides}, which at early times, can be written as:
\beq n_\phi=n_{_{\rm I}}\left(\frac{R}{R_{\rm I}}\right)^{-3}.
\label{nfe}\eeq
Inserting this into Eq. (\ref{Hini}), we obtain:
\beq H^2=H_{\rm I}^2\left(\frac{R}{R_{\rm
I}}\right)^{-3},\quad\mbox{with}\quad H^2_{\rm I}=
\frac{8\pi}{3M_P^2}\Delta_\phi m_\phi n_{_{\rm I}}. \label{Hear}
\eeq
At early times the last term of the left part side of Eq.
(\ref{fR}) can be safely ignored and it can be easily solved,
after substituting Eqs (\ref{nfe}) and (\ref{Hear}) in it, with
result:
\beq f_{\rm R}=\frac{2}{5}f_{\rm Rc}\ R_{\rm I}^{3/2}\left(
R^{5/2} -R_{\rm I}^{5/2}\right),\quad\mbox{with}\quad f_{\rm Rc}=
\Gamma_\phi\Delta_\phi H_{\rm I}^{-1} m_\phi n_{_{\rm I}}.
\label{fRsol} \eeq
Combining Eqs (\ref{fRsol}) and (\ref{rs}{\sf a}), we end up with
the following relation:
\beq T=\left(\frac{12}{\pi^2 g_{\rho*}}f_{\rm Rc}\right)^{1/4}
\left( \left( \frac{R}{R_{\rm I}}\right)^{-3/2} -\left(
\frac{R}{R_{\rm I}}\right)^{-4}   \right)^{1/4}\cdot
\label{TR}\eeq
The function $T(R/R_{\rm I})$ reaches at (see Fig. \ref{temp})
\beq \left(\frac{R}{R_{\rm I}}\right)_0=1.48,\quad\mbox{a maximum
value}\quad T_{\rm max}=0.767\left(\frac{12}{\pi^2g_{\rho*}}f_{\rm
Rc}\right)^{1/4}\cdot \label{Tmax}\eeq
%\left(\frac{8}{3}\right)^{2/5}=
\begin{floatingfigure}[l]
\hspace*{-.74in} \epsfig{file=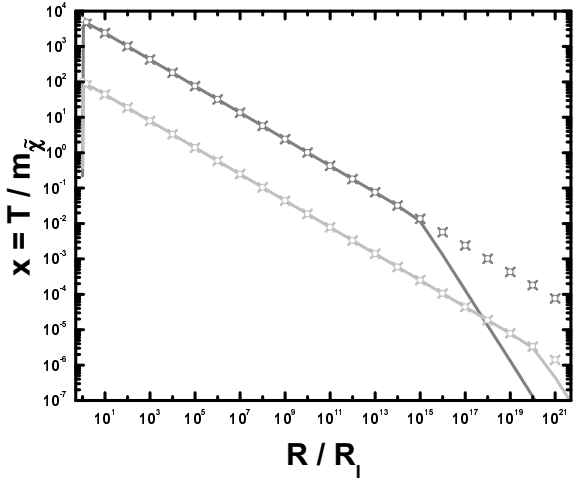,height=3.8in,angle=-90}
\hspace*{-.60in} \caption{\sl The evolution of
$x=T/m_{\tilde\chi}$ versus $R/R_{\rm I}$, derived from the
numerical solution of Eqs (\ref{nf})--(\ref{nx}) (solid lines),
from Eq. (\ref{TR}) (crosses) and from Eq. (\ref{RT}) (open
circles) for the parameters of Fig. 2-${\sf (a~[b])}$ (normal
[light] grey lines, crosses, circles).}
\label{temp}\end{floatingfigure}

~For $T<T_{\rm max}$ the second term in the second parenthesis in
Eq. (\ref{TR}) can be eliminated. From the resulting expression,
we can derive the following useful relations, which ``signalize''
the deviation from the standard RD evolution. Namely:

$\bullet$ Scale factor-temperature relation:
\beq g_{\rho*}^{2/3}\ R\ T^{8/3}=\left(\frac{12}{\pi^2}f_{\rm
Rc}\right)^{2/3}R_{\rm I}. \label{RT}\eeq
Recall that in the RD period is the quantity $g^{1/3}_{s*}(T)RT$
which remains constant.

$\bullet$ Expansion rate-temperature relation:
\beq H=\frac{\pi^2}{12}\ g_{\rho*}\ f^{-1}_{\rm Rc}\ H_{\rm I}\
T^4\,, ~~~~\label{Hphi}\eeq
which is obtained by solving Eq. (\ref{RT}) w.r.t $R$ and
inserting it into Eq. (\ref{Hear}). Note that during the RD phase,
$H$ is proportional to $T^2$ and not to $T^4$.

In Fig. 1, we show $x=T/m_{\tilde\chi}$ as a function of $R/R_{\rm
I}$ for the case of two representative examples which will be
analyzed in sec. \ref{thnonth}. The solid lines are obtained by
using Eq. (\ref{rs}{\sf a}) and the precise numerical solution of
Eqs (\ref{nf})--(\ref{nx}), whereas crosses [open circles] are
from the approximated Eq. (\ref{TR} [\ref{RT}]). Good agreement is
observed for $T_{\rm max}>T>T_{\rm RH}$ in both cases.

%In Fig. \ref{temp}, we depict the evolution of $T$ as a function
%of $R$ for the inputs indicated in Fig. 2. We observe a good
%agreement between the numerical solution and the result of Eq.
%(\ref{TR}) and a late time good agreement between them and the
%evolution described by Eq. (\ref{RT}).

\section{C{\ftn OLD} D{\ftn ARK} M{\ftn ATTER} A{\ftn BUNDANCE}}
\label{sec:neut}
\hspace{.562cm} The aim of this section is the calculation of
$\Omega_{\tilde\chi}h^2$ based on the already obtained
semi-analytical expressions. We assume that $\tilde\chi$'s are
non-relativistic around the `critical' temperatures $T_\ast$ or
$T_{\rm F}$ (see below) and never in chemical equilibrium after
the completion of reheating (however, see sec. \ref{omega}). The
relevant Boltzmann equation is properly re-formulated in sec.
\ref{boltzn}. Then two cases are investigated: $\tilde\chi$'s do
(sec. \ref{sec:eq}) or do not reach (sec. \ref{sec:noneq})
chemical equilibrium with plasma. The condition which
discriminates the two possibilities is specified in sec.
\ref{sec:noneq}.

\subsection{R{\ssz EFORMULATION OF THE}
B{\ssz OLTZMANN} E{\ssz QUATION}}\label{boltzn}

\hspace{.562cm} Our first step is the re-expression of
Eq.~(\ref{nx}) in terms of the variables $Y^{\rm [eq]}=n^{\rm
[eq]}_{\tilde\chi}/s^{8/3}$ in order to absorb the dilution term.
Armed with Eqs (\ref{RT}) and (\ref{Hphi}), we are able now to
convert the derivatives w.r.t $t$, to derivatives w.r.t
$x=T/m_{\tilde\chi}$. In fact, supposing that $g_{s*}(T)$ and
$g_{\rho *}(T)$ vary very little from their values at $T_{\rm RH}$
and differentiating the constant quantity $Rs^{8/9}$ (see Eq.
(\ref{RT}) and (\ref{rs}{\sf b})) w.r.t $t$, we obtain:
\begin{equation}
 -8\dot s= 9Hs \Rightarrow \dot x^{-1}=-8s^\prime/9Hs,
\label{eq:entropycons}
\end{equation}
where prime means derivative w.r.t $x$. Hence, Eq.~(\ref{nx}) can
be rewritten in terms of the new variables as:
\begin{equation} \label{BE3}
\dot Y = -\langle \sigma v \rangle\> \left( Y^2 - {Y^{\rm eq}}^2
\right) s^{8/3}+\Gamma_\phi N_{\tilde\chi} n_\phi s^{-11/3}.
\end{equation}
Replacing $R$ by $x$ in Eq. (\ref{nfe}), through Eq. (\ref{RT}),
we arrive at:
\beq \label{nfee} n_\phi(x)=n_{_{\rm I}}(x/x_{_{\rm I}})^{8}, \eeq
with $x_{_{\rm I}}=T_{\rm I}/m_{\tilde\chi}$ and $T_{\rm I}$ the
temperature corresponding to $R_{\rm I}$ derived from Eq.
(\ref{RT}). Differentiating Y w.r.t $x$ and replacing $\dot
x^{-1}$ by Eq. (\ref{eq:entropycons}), we obtain:
\beq \label{BE4}
%9sH Y^{\prime} =-8ds/dTm_{\tilde\chi}\dot Y
Y^{\prime}=\dot Y \dot x^{-1} \Rightarrow Y^{\prime}
=-\frac{8s^\prime}{9sH}m_{\tilde\chi}\dot Y. \eeq
Substituting Eqs (\ref{rs}{\sf b}), (\ref{Hphi}) and (\ref{BE3})
in Eq. (\ref{BE4}), this reads:
\\[-3mm]
\beq Y^{\prime}=y_{_n}\left(Y^2-Y^{\rm eq2}\right)
x^3-y_{_N}x^{-5},\label{BEf}\eeq\\[-12mm]
\bea \mbox{where}\quad && y_{_n}(x)=\frac{64}{45}
\left(\frac{2\pi^2}{45}\right)^{5/3} H_{\rm I}^{-1} f_{\rm Rc}\
m^4_{\tilde\chi}\ g_{\rm ol}\ \langle\sigma v\rangle \label{Yn}
\\ \mbox{and}\quad &&
y_{_N}(x)=\frac{64}{45}\left(\frac{2\pi^2}{45}\right)^{-11/3}
H_{\rm I}^{-1}\ N_{\tilde\chi}\ \Gamma_\phi\ f_{\rm Rc}\ n_{_{\rm
I}}\ x_{_{\rm I}}^{-8} m^{-12}_{\tilde\chi}\bar g_{\rm ol}.
\label{YN} \eea
The shorthand $g_{\rm ol}=g_{\rho*}^{-1/2}\ g_{s*}^{5/3}\
g_{*}^{1/2}$ and $\bar g_{\rm ol}=g_{\rho*}^{-1/2}\
g_{s*}^{-11/3}\ g_{*}^{1/2}$ has been used, with \cite{edjo}:
\begin{equation}\nonumber
g_{*}^{1/2} = \frac{g_{s*}}{\sqrt{g_{\rho*}}} \left( 1+\frac{x
g^\prime_{s*}}{3g_{s*}} \right)\cdot
\end{equation}

%\begin{equation}
%g_{\rm ol}^{[\prime]}=g_{\rho*}^{-1/2}\ g_{s*}^{5/3}\
%[g_{s*}^{-11/3}]\ g_{*}^{1/2},\quad\mbox{with \cite{edjo}}\quad
%g_{*}^{1/2} = \frac{g_{s*}}{\sqrt{g_{\rho*}}} \left(
%1+\frac{T}{3g_{s*}} \frac{d g_{s*}}{dT} \right)\cdot
%\end{equation}
%

\subsection{{\ssz NON}-E{\ssz QUILIBRIUM} P{\ssz RODUCTION}}
\label{sec:noneq}

\hspace{.562cm} In this case, $Y\ll Y^{\rm eq}$ and so,
$Y^2-Y^{\rm eq2}\simeq -Y^{\rm eq2}$. Inserting this into
Eq.~(\ref{BEf}) and integrating from $x_{_{\rm I}}$ down to
$x_{_{\rm RH}}=T_{\rm RH}/m_{\tilde\chi}$, we arrive at
$Y(x_{_{\rm RH}})=Y_{\rm RH}$:
\\[-3mm]
\beq Y_{\rm RH}=Y_{n}(x_{_{\rm RH}})+Y_{N}(x_{_{\rm RH}}),
\label{nonBEsol}\eeq\\[-12mm]
\bea \label{BEsolneq1} \mbox{with}\quad && Y_{n}(x)= \frac{64
}{45}\frac{g^2}{(2\pi)^3}\left(\frac{2\pi^2}{45}\right)^{-11/3}
H_{\rm I}^{-1}\ f_{\rm Rc}\ J_n(x)\ m_{\tilde\chi}^{-6}\\
\label{BEsolneq2} \mbox{and}\quad &&
Y_{N}(x)=\frac{64}{45}\left(\frac{2\pi^2}{45}\right)^{-11/3}H_{\rm
I}^{-1}\ N_{\tilde\chi}\ \Gamma_\phi\  J_{N}(x)\ f_{\rm Rc}\
m_{\tilde\chi}^{-12}\ n_{_{\rm I}}\ x_{_{\rm I}}^{-8}\ , \eea
where we have defined the quantities ($x_{_{\rm I}}$ can be
approximated by infinity, since  $x_{_{\rm I}}\gg x_{_{\rm RH}}$):
\beq J_n(x^\prime)= \int_{x^\prime}^{\infty}dx\ x^{-10} e^{-2/x}
P^2_2(1/x) g_{s*}^{-16/3} g_{\rm ol} \langle\sigma v\rangle\
\quad\mbox{and}\quad J_N(x^\prime)= \int_{x^\prime}^{\infty}dx\
x^{-5} \bar g_{\rm ol}\ . \label{jstar} \eeq

The maximum $\tilde\chi$ particles production takes place at
$x_\ast=T_\ast/m_{\tilde\chi}$ (=0.212, for constant
$\langle\sigma v\rangle$), where the integrand of $J_{n}$ reaches
its maximum \cite{riotto}. Therefore, the condition which
discriminates the EP from the non-EP of $\tilde\chi$'s:
\beq \label{cond} Y_{n}(x_\ast)+Y_{N}(x_\ast)-Y^{\rm eq}(x_\ast)
\left\{\matrix{
%\begin{array}{rl}
< 0\hfill ,& \mbox{non-EP} \hfill \cr
\geq 0 \hfill,& \mbox{EP} \hfill \cr}
%\end{array}
\right. \eeq

\subsection{E{\ssz QUILIBRIUM} P{\ssz RODUCTION}}\label{sec:eq}

\hspace{.562cm} In this case, we introduce the notion of freeze
out temperature, $T_{\rm F}=x_{_{\rm F}}m_{\tilde\chi}$
\cite{kolb, gelmini}, which assists us to study Eq.~(\ref{BEf}) in
the two extreme regimes:

$\bullet$ At very early times, when $x\gg x_{_{\rm F}}$,
$\tilde\chi$'s are very close to equilibrium. So, it is more
convenient to rewrite Eq.~(\ref{BEf}) in terms of the variable
$\Delta(x)=Y(x)-Y^{\rm eq}(x)$ as follows:
\beq \label{deltaBE} \Delta^{\prime}=-{Y^{\rm eq}}^{\prime}+y_{_n}
\Delta\left(\Delta+2Y^{\rm eq}\right)\ x^3-y_{_N}\ x^{-5}. \eeq
The freeze-out temperature can be defined by
\beq \Delta(x_{_{\rm F}})=\delta_{\rm F}\>Y^{\rm eq}(x_{_{\rm F}})
\Rightarrow \Delta(x_{_{\rm F}})\Big(\Delta(x_{_{\rm F}})+2Y^{\rm
eq}(x_{_{\rm F}})\Big)=\delta_{\rm F}(\delta_{\rm F}+2)\ Y^{\rm
eq2}(x_{_{\rm F}}), \label{Tf} \eeq
where $\delta_{\rm F}$ is a constant of order one determined by
comparing the exact numerical solution of Eq.~(\ref{BEf}) with the
approximate under consideration one. Inserting Eqs~(\ref{Tf}) into
Eq.~(\ref{deltaBE}) and neglecting $\Delta^{\prime}$ (since
$\Delta^{\prime}\ll{Y^{\rm eq}}^{\prime}$), we obtain the
following equation, which can be solved w.r.t $x_{_{\rm F}}$
iteratively:
\bea && \Big(\ln Y^{\rm eq}(x_{_{\rm F}})\Big)^\prime =y_{_{n{\rm
F}}}\delta_{\rm F} (\delta_{\rm F}+2) Y^{\rm eq}(x_{_{\rm
F}})x_{_{\rm F}}^3- y_{_{N{\rm F}}} x_{_{\rm F}}^{-5}/Y^{\rm
eq}(x_{_{\rm F}}), \quad \mbox{with} \label{xf}
\\ \label{xfa}&& y_{_{N[n]\rm F}}= y_{_{N[n]}}(x_{_{\rm F}})
\quad\mbox{and}\quad\Big(\ln Y^{\rm
eq}(x)\Big)^{\prime}=\frac{1}{x^2}-\frac{13}{2x}-
\frac{8g^{\prime}_{s*}}{3g_{s*}}-\frac{P^\prime_2(1/x)}{x^2P_2(1/x)}\cdot
\eea

$\bullet$ At late times, when $x\ll x_{_{\rm F}}$, $Y\gg Y^{\rm
eq}$ and so, $Y^2-Y^{\rm eq2}\simeq Y^2$. Inserting this into
Eq.~(\ref{BEf}), the value of $Y$ at $T_{\rm RH}$, $Y_{\rm
RH}=Y(x_{_{\rm RH}})$ can be found by solving the resulting
differential equation. We distinguish the cases:

\hspace{.6cm} {\bf i.} For $N_{\tilde\chi}=0$, the integration
from $x_{_{\rm F}}$ down to $x_{_{\rm RH}}$ can be made trivially,
with result:
\begin{equation} \label{BEsol}
Y_{\rm RH} = \left(\frac{1}{Y_{\rm
F}}+\frac{8}{15\pi}\left(\frac{2\pi^2}{45}\right)^{5/3} M_{\rm
P}^2\>\Gamma_\phi\>m_{\tilde\chi}^4\> J_{\rm F} \right)^{-1},
\end{equation}
where we have defined the quantities:
\bea \label{sigmaeff3} && J_{\rm F}= \int_{x_{_{\rm
RH}}}^{x_{_{\rm F}}} dx\ x^3 g_{\rm ol}\ \langle\sigma v\rangle
~~\mbox{and}~~Y_{\rm F} =(\delta_{\rm F} +1)\> Y^{\rm eq}(x_{_{\rm
F}}),\\
\mbox{with} \quad &&  \label{defYF} Y^{\rm
eq}(x)=\frac{g}{(2\pi)^{3/2}}\left(\frac{2\pi^2}{45}\right)^{-8/3}
g_{s*}^{-8/3}\> m_{\tilde\chi}^{-5}\>x^{-13/2}\>e^{-1/x}\
P_2(1/x). \eea
The choice $\delta_{\rm F}=1.0\mp0.2$ provides the best agreement
with the precise numerical solution of Eq.~(\ref{BEf}), without to
cause dramatic instabilities.

\hspace{.5cm} {\bf ii.} For $N_{\tilde\chi}\neq0$, the integration
of the resulting equation can be realized numerically. However,
fixing $\langle\sigma v\rangle$ and $g$'s to their values at
$x_{_{\rm F}}$, an analytic formula can be derived for this case,
also (note that $x_{_{\rm RH}}<x_{_{\rm F}}$):
\begin{equation} \label{NBEsol}
Y_{\rm RH} = \frac{1}{y_{_{n \rm F}}x^4_{_{\rm
RH}}}\left(-2+y_{_{\rm F}}\tan \Big(\tan^{-1}\frac{2+y_{_{n\rm F}
}x_{_{\rm F}}Y_{\rm F}}{y_{_{\rm F}}}+y_{_{\rm F}}\ln
\frac{x_{_{\rm RH}}}{x_{_{\rm F}}}\Big) \right),\eeq
where $Y_{\rm F}$ is given by Eq. (\ref{defYF}) and the quantity
$y_{_{\rm F}}=\sqrt{-4-y_{_{n \rm F}}y_{_{N \rm F}}}$ has been
defined.

\subsection{T{\ssz HE} C{\ssz URRENT} A{\ssz BUNDANCE}}
\label{omega}

\hspace{.562cm} Our final aim is the calculation of the current
$\tilde\chi$ relic density, which is based on the well known
formula \cite{gelmini, edjo}:
\begin{equation}
\label{om1} \Omega_{\tilde\chi}=\rho_{\tilde\chi}^0/\rho_{\rm
c}^0= m_{\tilde\chi} s_0 Y_0/\rho_{\rm c}^0,~~{\rm
where}~~Y_0=n_{\tilde\chi}^0/s_0,
\end{equation}
$\rho_{\tilde\chi}^0=m_{\tilde\chi} n_{\tilde\chi}^0$ is the
current $\tilde\chi$ energy density and $s_{0}~[\rho_{\rm c}^0]$
is the entropy [critical energy] density today. With a background
radiation temperature of $T_0=2.726~^0$K, we arrived at the final
result:
\begin{equation} \label{eq:omegah2}
\Omega_{\tilde\chi}h^2 = 2.741 \times 10^8\ Y_0\
m_{\tilde\chi}/\mbox{GeV}.
%\frac{m_{\tilde\chi}}{\mbox{GeV}}\cdot
\end{equation}

For the numerical program, $Y_0$ is determined at a value $R_{\rm
f}$ large enough so as, $Y_0$ is stabilized to a constant value
(with $g$'s fixed to their values at $T_{\rm RH}$). For the
semi-analytical calculation, we assume that there is no entropy
production for $T<T_{\rm RH}$. Then, it is convenient to single
out the cases:

\hspace{.6cm} {\bf i.} For $N_{\tilde\chi}=0$, there is no
${\tilde\chi}$ production for $x<x_{_{\rm RH}}$. Therefore, the
present value of $Y$ can be estimated reliably at $x_{_{\rm RH}}$
with $Y_0= Y_{\rm RH}\ s^{5/3}(x_{_{\rm RH}})$ (in accord with
Ref. \cite{riotto}).

\hspace{.5cm} {\bf ii.} For $N_{\tilde\chi}\neq0$, the residual
produced $\tilde\chi$'s (i.e., for $x<x_{_{\rm RH}}$) can
annihilate satisfying an equation analog to Eq. (\ref{nx}) but in
a RD background, any more, determined by Eq. (\ref{rs}{\sf a})
(similarly to the case of Ref. \cite{fujii}):
\beq \dot n_{\tilde\chi}+3H_{\rm RD}n_{\tilde\chi}+\langle \sigma
v \rangle \left( n_{\tilde\chi}^2 - n_{\tilde\chi}^{\rm
eq2}\right)-\Gamma_\phi N_{\tilde\chi} n_\phi=0
,~~\mbox{with}\quad H^2_{\rm RD}=\frac{8\pi}{3M^2_{\rm P
}}\rho_{_{\rm R}}. \label{nxRD} \eeq
The equilibrium distribution $n_{\tilde\chi}^{\rm eq}$ can be
safely neglected since, as it turns out, is in any case
numerically irrelevant. Using the variable $Y_{\rm
RD}=n_{\tilde\chi}/s$ and the entropy conservation law in the RD
era, Eq. (\ref{nxRD}) can be rewritten as:
\begin{equation} \label{BERD}
Y^\prime_{\rm RD} = \frac{s^\prime}{3H_{\rm RD}}\left(Y^2_{\rm RD}
\langle\sigma v\rangle -\Gamma_\phi N_{\tilde\chi} n_\phi
s^{-2}\right),\quad\mbox{with \cite{edjo}}\quad
\frac{s^\prime}{3H_{\rm RD}}=\sqrt{\frac{\pi g_*}{45}}\
\frac{M_{\rm P}}{m_{\tilde\chi}}\cdot\end{equation}
The post-late times, i.e. for $x<x_{_{\rm RH}}$, evolution of
$n_\phi$ can be approximated, using the exact solution of Eq.
(\ref{nf}) \cite{turner}, the entropy conservation law and the
time-temperature relation \cite{lazarides} in a RD era, as follows
(the ratio $x_{_{\rm RH}}/x_{_{\rm I}}$ can be safely ignored):
\beq \label{nflt} n_\phi(x)=n_{_{\rm RH}}\left(x/x_{_{\rm
RH}}\right)^3 g_{s*}(x)/g_{s*}(x_{_{\rm RH}})  e^{-x^2_{\rm
RH}/x^{2}},\eeq
where $n_{_{\rm RH}}=n_\phi(x_{_{\rm RH}})$ is given by Eq.
(\ref{nfe}). Inserting Eqs (\ref{nflt}) and (\ref{rs}{\sf b}) into
Eq. (\ref{BERD}), this can be cast in the following final form:
\beq Y_{\rm RD}^{\prime}=y_{_{{\rm RD}n}}Y_{\rm RD}^2-y_{_{{\rm
RD}N}}\ e^{-x^2_{\rm RH}/x^{2}} x^{-3},\label{BEfRD}\eeq\\[-12mm]
\bea \mbox{where}\quad && y_{_{{\rm
RD}n}}(x)=\sqrt{\frac{\pi}{45}} M_{\rm P} \ m_{\tilde\chi}\
g_*^{1/2}\ \langle\sigma v\rangle \label{YRDn} \\ \mbox{and}\quad
&& y_{_{{\rm
RD}N}}(x)=\sqrt{\frac{\pi}{45}}\left(\frac{2\pi^2}{45}\right)^{-2}
M_{\rm P} N_{\tilde\chi}\ \Gamma_\phi\ n_{_{\rm RH}} x_{_{\rm
RH}}^{-3}\ m^{-5}_{\tilde\chi} g_*^{1/2} g_{s*}^{-1}
g_{s*}^{-1}(x_{_{\rm RH}}). \label{YRDN} \eea
$Y_0$ can be obtained, by solving numerically Eq. (\ref{BEfRD})
from $x_{_{\rm RH}}$ down to $0$, with initial condition $Y_{\rm
RD}(x_{_{\rm RH}})=Y(x_{_{\rm RH}})s^{5/3}(x_{_{\rm RH}})$,
$Y(x_{_{\rm RH}})$ being derived from Eq. (\ref{NBEsol}) or
(\ref{nonBEsol}). Even in the exceptional case, (see sec.
\ref{numan}) where $\tilde\chi$ decouples  a little after the
completion of reheating (i.e., for $x<x_{_{\rm RH}}$), Eq.
(\ref{BEfRD}) can be applied with $Y_{\rm RD}(x_{_{\rm RH}})=0$,
giving reliable results. When the first [second] term in the right
hand side of Eq. (\ref{BEfRD}) evaluated at $x_{_{\rm RH}}$
dominates over the second [first] one, an analytical solution of
Eq. (\ref{BEfRD}) can be easily derived, which works frequently
well, as we will see in sec. \ref{numan}:
\beq \label{BEfan} {\sf (a)}~~ Y_0=\frac{Y_{\rm RD}(x_{_{\rm
RH}})}{1+y_{_{{\rm RD}n}}(x_{_{\rm RH}}) Y_{\rm RD}(x_{_{\rm
RH}})x_{_{\rm RH}}} \quad\Big[{\sf (b)}~~ Y_0=Y_{\rm RD}(x_{_{\rm
RH}})+\frac{y_{_{{\rm RD}N}}(x_{_{\rm RH}})}{2 e x_{_{\rm RH}}}
\Big],\eeq
where  $\langle\sigma v\rangle$ and $g$'s have been fixed at their
values at $x_{_{\rm RH}}$. Practically, since an unavoidable
discrepancy enters between the input $x_{_{\rm RH}}$ and the
output $x_{\rm rh}=T_{\rm rh}/m_{\tilde\chi}$ (i.e., the solution
of Eq. (\ref{Trhp})), $x_{_{\rm RH}}$ in Eqs (\ref{nflt}),
(\ref{BEfRD}), (\ref{YRDN}) and (\ref{BEfan}) is to be replaced by
a matching scale $x_\delta= x_{_{\rm RH}}/\delta_{\rm RH}$ with
$\delta_{\rm RH}=0.9\pm0.3$, in order for the solution of Eq.
(\ref{BEfRD}) to match better the solution of the system of Eqs
(\ref{nf})--(\ref{nx}). Although we did not succeed to achieve a
general analytical solution of Eq. (\ref{BEfRD}), we consider as a
significant development the derivation of a result for our problem
by solving numerically just one equation, instead of the whole
system above.

\section{A{\ftn PPLICATIONS}}\label{appl}

\hspace{.562cm} Our numerical investigation depends on the
parameters:
$$m_\phi,\ N_{\tilde\chi},\ T_{\rm RH},\ m_{\tilde\chi},\
\langle\sigma v\rangle\ .$$
Note that the numerical choice of $n_{_{\rm I}}$ turns out to be
irrelevant for the result of $\Omega_{\tilde\chi} h^2$. Just for
definiteness we determine it, through $m_\phi$ using Eq.
(\ref{Hear}) with $H_{\rm I}=m_\phi$, as in the simplest model of
chaotic inflation \cite{quevedo, chung}. $\Gamma_\phi$, or
equivalently $T_{\rm RH}$, and $m_\phi$ can be related through a
coefficient which includes an effective suppression scale of the
interaction of $\phi$, $M_{\rm eff}$ and the coupling of  $\phi$
and ${\tilde\chi}$, $\lambda$ \cite{moroi, shaaban}. However,
since we examine the problem from cosmological point of view, we
prefer to handle $m_\phi$ and $T_{\rm RH}$ as input parameters
without any further reference to $M_{\rm eff}$ and $\lambda$,
which are obviously particle physics model dependent. In our
scanning we take into account the following bounds:
\beq {\sf (a)}~~10^3~{\rm GeV}\leq
m_\phi\lesssim8\times10^{14}~{\rm GeV} ,\quad {\sf
(b)}~~N_{\tilde\chi}\leq 1\quad\mbox{and}\quad{\sf (c)}~~T_{\rm
RH}\geq0.001~{\rm GeV}\ . \label{para}\eeq
The lower bound of Eq. (\ref{para}{\sf a}) is just conventional,
whereas the upper bound is required, such as the decay products of
the inflaton are thermalized within a Hubble time, through $2\to3$
processes \cite{sarkar}. The later is crucial so that Eqs
(\ref{nf})--(\ref{nx}) are applicable. The bound of Eq.
(\ref{para}{\sf b}) comes from the arguments of the appendix of
Ref. \cite{moroi} and this of Eq. (\ref{para}{\sf c}) from the
requirement not to spoil the successful predictions of Big Bang
nucleosynthesis.

Model dependent is also the derivation of $\langle\sigma v\rangle$
from $m_{\tilde\chi}$ and the residual {\sc SUSY} spectrum. To
keep our presentation as general as possible, we decide to treat
$m_{\tilde\chi}$ and $\langle\sigma v\rangle$ as unrelated input
parameters, choosing plausible (from the viewpoint of  {\sc SUSY}
models) values for them. Namely, we focus our attention on the
ranges:
\beq {\sf (a)}~~200~{\rm GeV} \leq m_{\tilde\chi}\leq500~{\rm GeV}
\quad\mbox{and}\quad {\sf (b)}~~10^{-12}~{\rm
GeV}^{-2}\leq\langle\sigma v\rangle\leq10^{-8}~{\rm GeV}^{-2}\ .
\label{parb}\eeq
The lower bound in Eq. (\ref{parb}{\sf a}) arises roughly from the
experimental constraints on the SUSY spectra (see, e.g., Fig. 23
of Ref. \cite{wmapl}), whereas the upper is imposed in order the
analyzed range to be possibly detectable in the future experiments
(see, e.g. Ref. \cite{munoz}). On the other hand, $\langle\sigma
v\rangle$ of the range of Eq. (\ref{parb}{\sf b}) can be naturally
produced in the context of SUSY models (see, e.g., Fig. 1 of Ref.
\cite{fornengo}). The lower value can be derived in the case of a
bino LSP or sleptonic coannihilations \cite{ellis2} whereas the
upper, in the case of a wino LSP \cite{fujii} or coannihilations
with squarks or gauginos or of $A$-pole effect \cite{santoso1}.
Note that the $x$-dependence (which turns out to be important only
for EP) in the case of a bino LSP  can be reliably absorbed, by
fixing $x$ in $\langle\sigma v\rangle$ to $x_{_{\rm F}}$ (e.g., if
we had posed $\langle\sigma v\rangle=10^{-10}x~{\rm GeV}^{-2}$ for
the residual parameters of Figs 1-(${\sf a_1, a_2}$), we would
have obtained $\Omega_{\tilde\chi}h^2\simeq0.57$, which could,
also, be derived by imposing $\langle\sigma
v\rangle=10^{-10}/21~{\rm GeV}^{-2}$).

In sec. \ref{thnonth} we will illustrate the two kinds of
${\tilde\chi}$ production and in sec. \ref{numan} we will compare
the results of our numerical and semi-analytical
$\Omega_{\tilde\chi}h^2$ calculations. Finally, in sec. \ref{NTR}
we will present areas compatible with Eq. (\ref{cdmb}).

\subsection{E{\ssz QUILIBRIUM} V{\ssz ERSUS}
{\ssz NON}-E{\ssz QUILIBRIUM} P{\ssz RODUCTION}} \label{thnonth}

\begin{floatingtable}[!h]
\caption{\sl Input and output parameters for the two examples
illustrated in Figs 1 and 2-{\sf (a, b)}.}
\begin{tabular}{|l||c|c|} \hline
{\bf \nsz F\ssz IGURE\nsz}&{\bf \nsz 2-${\sf (a_1, a_2)}$ \nsz}&
{\bf \nsz 2-${\sf (b_1, b_2)}$ \nsz} \\ \hline\hline
\multicolumn{3}{|c|}{\bf \nsz I\ssz NPUT \nsz P\ssz ARAMETERS \nsz
}\\ \hline
\multicolumn{3}{|c|}{$m_{\phi[{\tilde\chi}]}=10^6~[350]~{\rm
GeV},\langle\sigma v\rangle=10^{-10}~{\rm GeV}^{-2}$}\\ \hline
$x_{_{\rm RH}}$ &{5/350}&{0.001/350}
\\
$N_{\tilde\chi}$ &{$1.4\times10^{-7}$}&{$1.2\times10^{-3}$}\\
\hline
$g_{\rho\ast}(x_{_{\rm RH}})$&{10.88}&{3.36}\\
$g_{s\ast}(x_{_{\rm RH}})$ &{10.85}&{3.91}\\ \hline \hline
\multicolumn{3}{|c|}{\bf \nsz O\ssz UTPUT \nsz P\ssz ARAMETERS
\nsz }\\ \hline
\multicolumn{3}{|c|}{\bf \boldmath $\tilde\chi$-\nsz D\ssz
ECOUPLING}\\ \hline
%{\bf P\ssz RODUCTION \nsz:}&\multicolumn{2}{|c||}{E\ssz QUILIBRIUM
%\nsz} &\multicolumn{2}{|c||}{\nsz NON-E\ssz QUILIBRIUM \nsz}\\
%\hline
%
$(Y_n+Y_N)(x_{\ast})$&$1.4\times10^{-10}$&$2.0\times10^{-18}$\\
$Y^{\rm eq}(x_{\ast})$&$2.8\times10^{-16}$&$2.8\times10^{-16}$\\
\hline
\multicolumn{3}{|c|}{\bf \nsz P\ssz RECISE \nsz R\ssz EHEATING
\nsz}\\ \hline
$x_{\rm max}$&4378.2&80.4 \\
$x_{\rm rh}$&4.53/350&0.0009/350\\ \hline
$\rho_\phi(x_{\rm rh})~({\rm GeV}^4)$&$11826.3$
&$2.3\times10^{-12}$\\ \hline
$(R/R_{\rm I})_{\rm rh}$&$8.06\times10^{14}$&$1.38\times10^{20}$
\\ \hline
\end{tabular}
\end{floatingtable}

The operation of the two fundamental phenomena described in our
wo-rk (i.e., the $\tilde\chi$ EP [non-EP]  and the reheating) are
instructively displayed in Figs 2-${\sf (a~[b])}$. Namely, we
depict in Figs 2-${\sf (a_1, b_1)}$,  $\rho^{[\rm
eq]}_{\tilde\chi}/s$ (solid [dotted] lines) and in Figs 2-${\sf
(a_2, b_2)}$, the $\phi$ [radiation] energy density
$\rho_{\phi[{_{\rm R}}]}$ (dashed [dot dashed] lines) versus $x$
(we prefer $\rho^{[\rm eq]}_{\tilde\chi}/s$ instead of $Y_{\rm
(RD) }^{\rm [eq]}$, since it offers a unified description of the
evolution before and after $x_{_{\rm RH}}$). The needed for our
calculation inputs and some key-outputs are listed in Table 1.

In both cases, we used the same $m_\phi$ and we fixed
$m_{\tilde\chi}$ and $\langle\sigma v\rangle$ in the middle of the
ranges of Eq. (\ref{parb}), which correspond to
$\Omega_{\tilde\chi}h^2=1.87$ with $x_{_{\rm F}}=22.25/350$ for
the standard paradigm (see sec. \ref{NTR}). By adjusting $T_{\rm
RH}$ and $N_{\tilde\chi}$, we achieve EP or non-EP, extracting the
central value of $\Omega_{\tilde\chi}h^2$ in Eq. (\ref{cdmba}).
The distinguish is realized by explicitly applying the criterion
of Eq. (\ref{cond}) for $x_\ast=74.2/350$ (see Table 1) and is
illustrated in Figs 2-${\sf (a_1, b_1)}$. Higher $T_{\rm RH}$, and
consequently (see sec. \ref{numan}), lower $N_{\tilde\chi}$ are
required for EP. In this case, the solution of Eq. (\ref{xf}) is
$x_{_{\rm F}}=20.32/350$. The choice $\delta_{\rm RH}=1.15~[0.68]$
allows us to reach the numerical solution of Eqs
(\ref{nf})--(\ref{nx}), by solving Eq. (\ref{BEfRD}) or employing
Eq. (\ref{BEfan}${\sf b}$). \\ $~~~~~~~$ Let us, now, stress on
the crucial hierarchies encountered in our examples (see Table 1)
and explain their implications, in conjunction with the analyses
of Refs \cite{riotto, dreesa, notari}:

%The use of our definition in Eq. (\ref{GTrh}) allows us to
%approach quite successfully $x_{\rm rh}$ by $x_{_{\rm RH}}$ (in
%contrast with Ref. \cite{riotto}).
% causing an equally significant instability on the
%$\Omega_{\tilde\chi}h^2$ calculation, too.

%%%%%%%%%%%%%%%%%%%%%%%%%%%%%%%%%%%%%%%%%%%%%%%%%%%%%%%%%%%%%%%%%%%%
\begin{figure}[t]\vspace*{-.19in}
\hspace*{-.71in}
\begin{minipage}{8in}
\epsfig{file=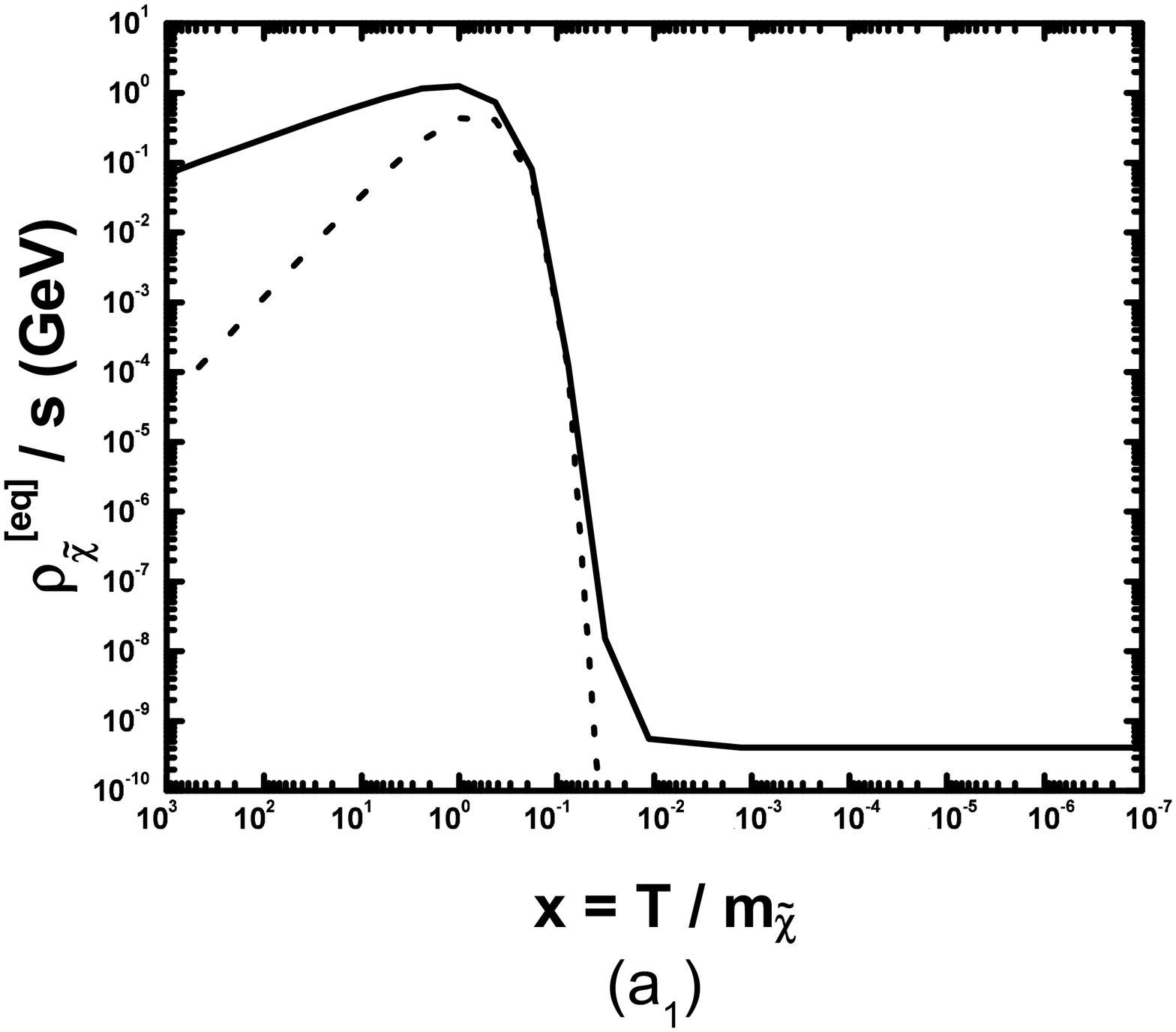,height=3.8in,angle=-90} \hspace*{-1.37 cm}
\epsfig{file=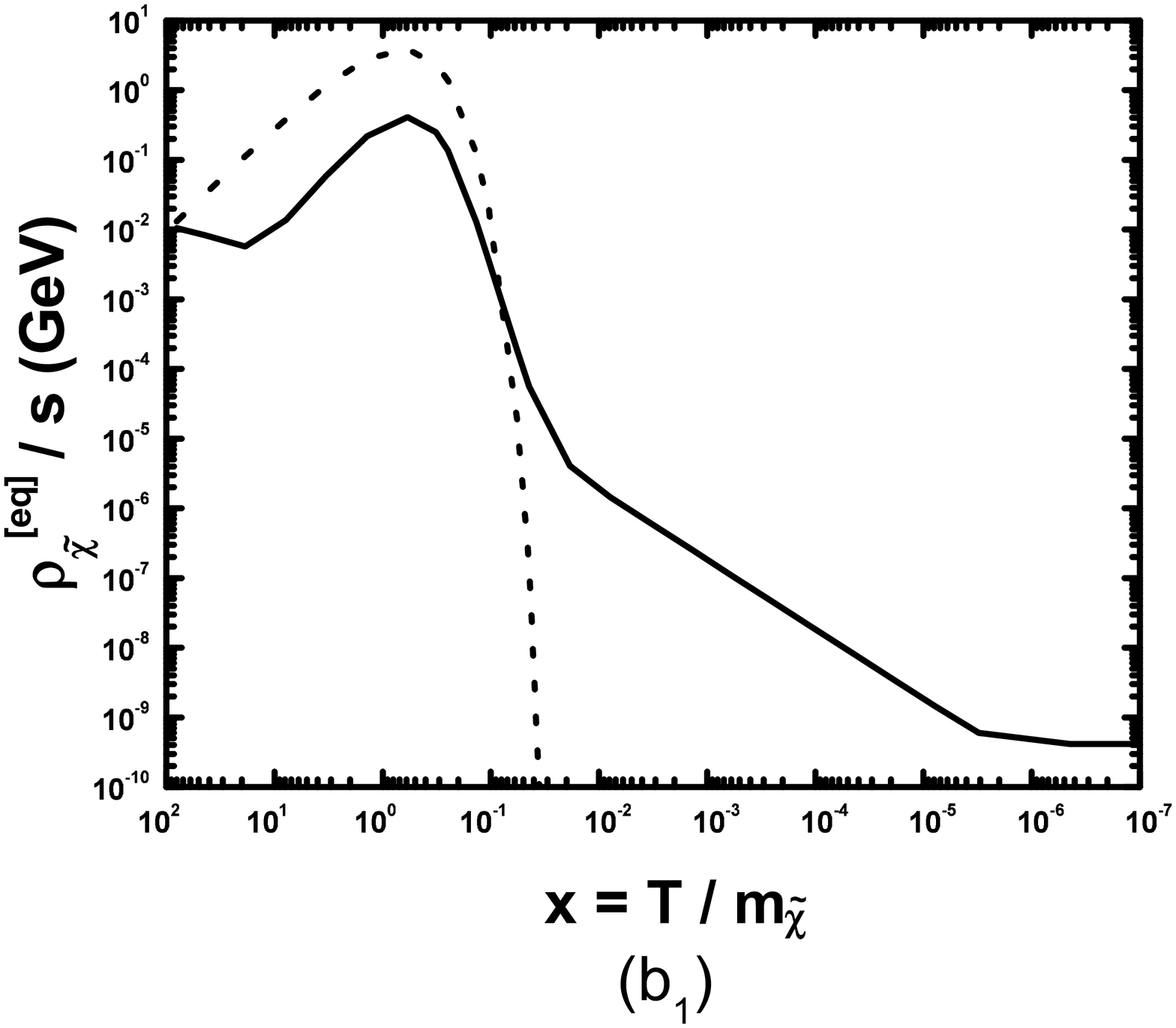,height=3.8in,angle=-90} \hfill
\end{minipage}\vspace*{-.01in}
\hfill\hspace*{-.71in}
\begin{minipage}{8in}
\epsfig{file=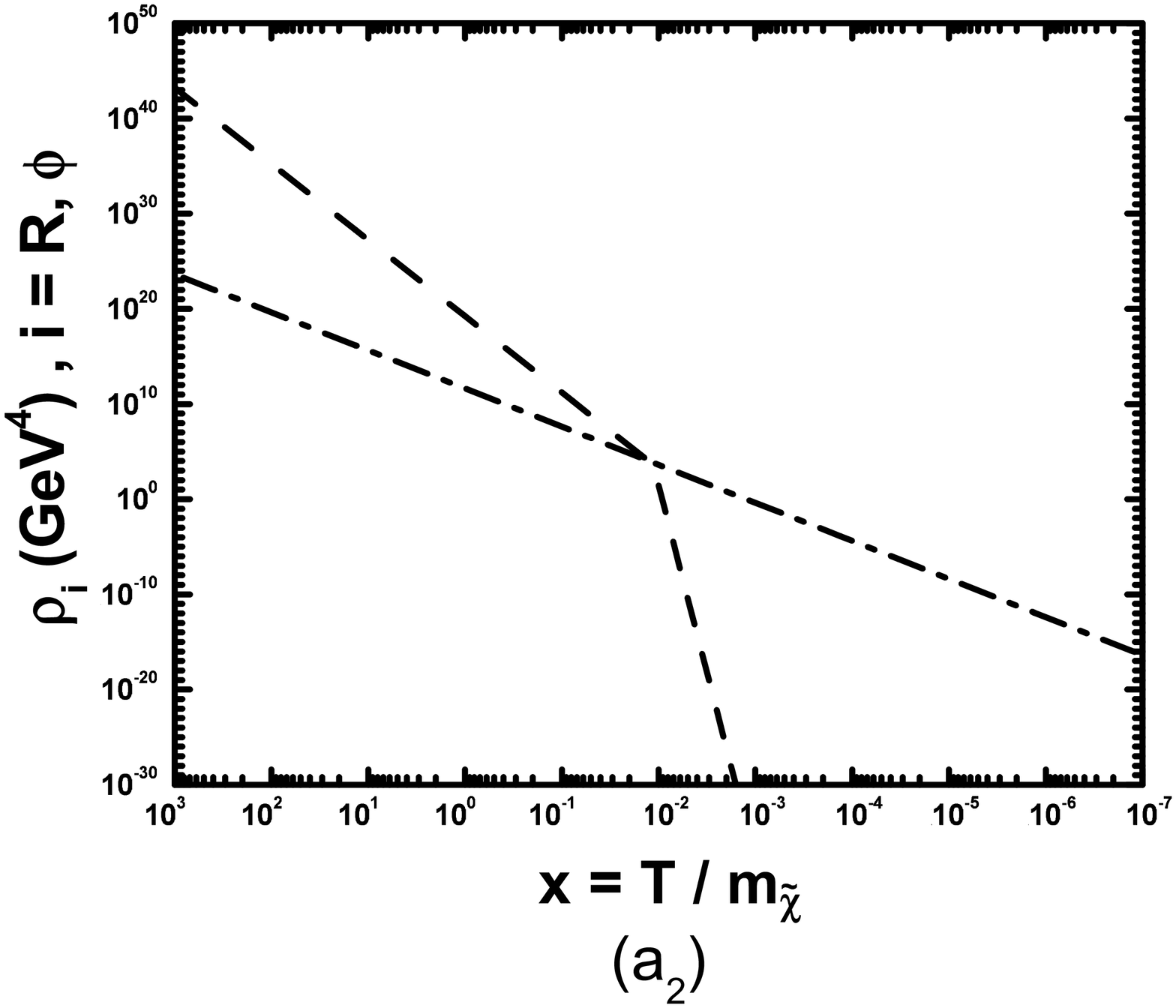,height=3.8in,angle=-90} \hspace*{-1.37 cm}
\epsfig{file=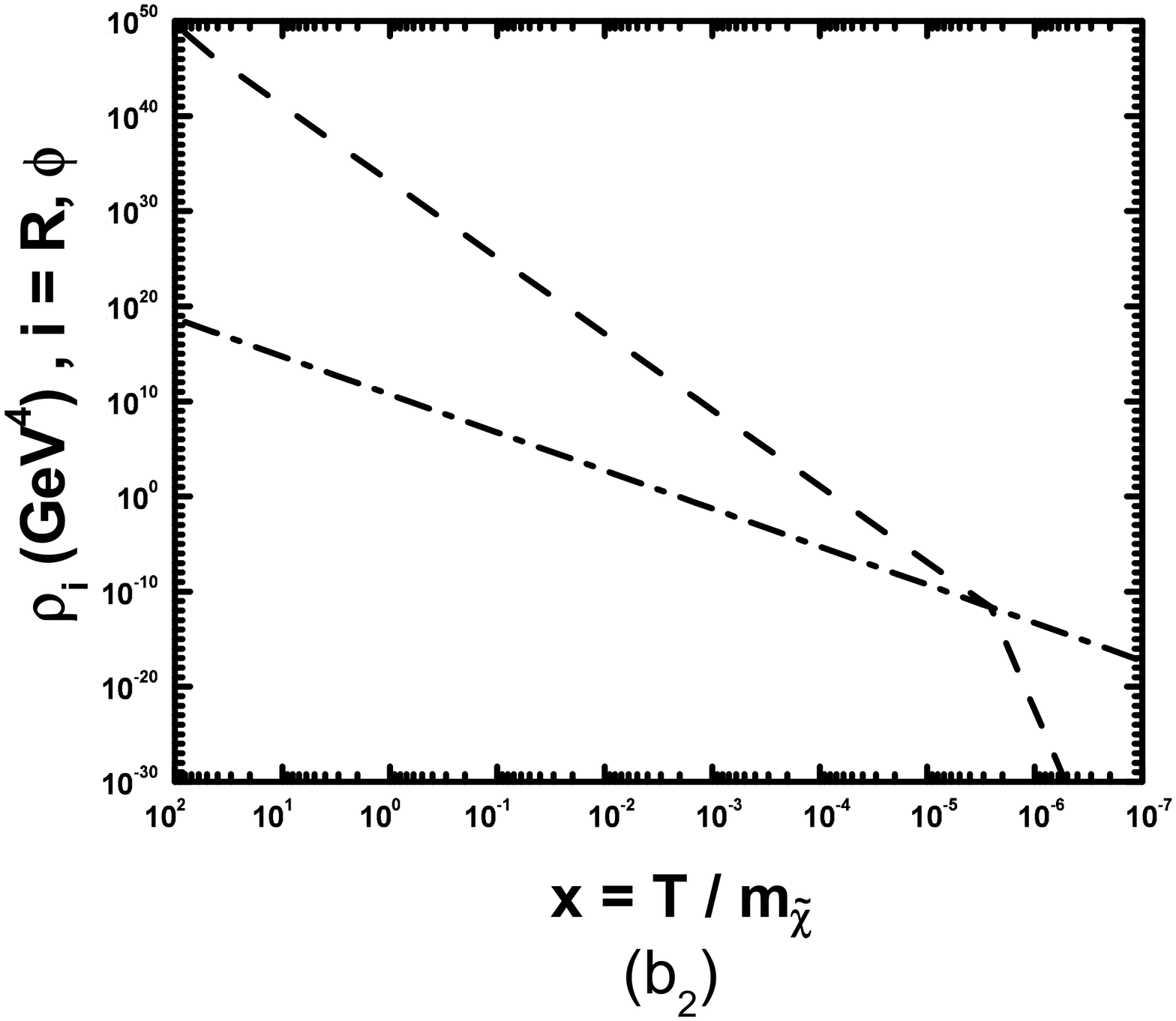,height=3.8in,angle=-90} \hfill
\end{minipage}
\hfill \caption[]{\sl ${\sf (a_1, b_1)}$ The quantity $\rho^{[\rm
eq]}_{\tilde\chi}/s$ (solid [dotted] line) and ${\sf (a_2, b_2)}$
the $\phi$ [radiation] energy densities $\rho_{\phi[{_{\rm R}}]}$
(dashed [dot dashed] line) versus $x=T/m_{\tilde\chi}$. We take
$m_\phi=10^6~{\rm GeV}, m_{\tilde\chi}=350~{\rm GeV},
\langle\sigma v\rangle=10^{-10}~{\rm GeV}^{-2}$ and $T_{\rm
RH}=5~{\rm GeV},~N_{\tilde\chi}=1.4\times10^{-7}$ ${\sf (a_1,
a_2)}$ for EP or $T_{\rm RH}=0.001~{\rm GeV},~
N_{\tilde\chi}=1.2\times10^{-3}$ ${\sf (b_1, b_2)}$ for non-EP. In
both cases, we extract $\Omega_{\tilde\chi}h^2=0.115$.}
\label{fig:eqnoneq} \vspace*{-.11in}
\end{figure}
%%%%%%%%%%%%%%%%

$\bullet$ In both cases, $T_{\rm max}=x_{\rm max}m_{\tilde \chi}>
T_{\rm rh}=x_{\rm rh}m_{\tilde \chi}$, with $x_{\rm rh}$ being
derived by solving Eq. (\ref{Trhp}) and corresponding to
$(R/R_{\rm I})_{\rm rh}\gg (R/R_{\rm I})_0$ (see Fig. 1). Our
definition in Eq. (\ref{GTrh}) allows us to approach quite
successfully $x_{\rm rh}$ by $x_{_{\rm RH}}$ (in contrast with
Ref. \cite{riotto}). Our $T_{\rm max}\sim m^{1/4}_\phi
M^{1/4}_{\rm P}T^{1/2}_{\rm RH}$, in accord with Ref.
\cite{riotto}, turns out to be larger by about one order of
magnitude than this, estimated in Ref. \cite{dreesa}.

$\bullet$ In both cases, $T_{\rm max}>m_{\tilde\chi}>T_{\rm
RH}/x_{_{\rm F}(\ast)}$. This means that $\tilde\chi$ is first
relativistic and then becomes non-relativistic, and so, its
characterization as cold relic is self-consistent.

$\bullet$ In Fig 2-${\sf (a~[b])}$, we have $T_{\rm
max}\simeq[<]m_{\phi}$. So, our perturbative approach to the
$\phi$ decay is well defined \cite{dreesa} and possible effects of
plasma masses are certainly negligible \cite{notari}.

Let us finally note that hard-hard and hard-soft contributions to
$\Omega_{\tilde\chi}h^2$ \cite{dreesa} are not included in our
approach. For the ranges of Eqs (\ref{para}) and (\ref{parb}), the
hard-hard contributions are certainly negligible, whereas the
hard-soft are mostly sub-dominant, since $m_\phi\gg
m_{\tilde\chi}$.

%However, a detail comparison requires deeper investigation.

\subsection{N{\ssz UMERICAL} V{\ssz ERSUS} S{\ssz
EMI}-A{\ssz NALYTICAL} R{\ssz ESULTS}}
\label{numan}

\begin{table}[!h]
\begin{center}
\begin{tabular}{|c|c|c||c|c|}
\hline {\bf  F\ssz IG. \nsz} & \multicolumn{2}{|c||}{\bf \nsz
R\ssz ANGES OF \nsz \boldmath $x$-A\ssz XIS \nsz P\ssz ARAMETERS
\nsz}& {\bf \boldmath $\tilde\chi$-P\ssz RO-}& {\boldmath
$\delta_{\rm RH}$}\\ \cline{2-3}
& {\bf $m_{\tilde\chi}=200~{\rm GeV}$}&{\bf
$m_{\tilde\chi}=500~{\rm GeV}$} &{\bf\ssz DUCTION}&
\\ \hline \hline
\ref{om}-${\sf (a_1)}$&$(0.001-0.07)~{\rm GeV}$&$(0.001-0.12)~{\rm
GeV}$&non-EP & $0.8$ \\
&$(0.07-2.5)~{\rm GeV}$&$(0.12-4)~{\rm GeV}$ & EP & $1.1$
\\\hline
\ref{om}-${\sf (b_1)}$ &\multicolumn{2}{|c||}{$(10^3-10^7)~{\rm
GeV}$} & non-EP& $0.8$\\ \hline
\ref{om}-${\sf (c_1)}$ &\multicolumn{2}{|c||}{$10^{-7}-10^{-3}$} &
non-EP& $0.8$
\\ \hline
\ref{om}-${\sf (a_2)}$ &$(0.001-0.015)~{\rm
GeV}$&$(0.0015-0.0052)~{\rm GeV}$&EP & $0.75$ \\
&$(0.4-100)~{\rm GeV}$&$(1-100)~{\rm GeV}$ & EP & $0.8$ \\\hline
\ref{om}-${\sf (b_2)}$ &\multicolumn{2}{|c||}{$(10^3-10^9)~{\rm
GeV}$} & EP& $0.85$\\
&\multicolumn{2}{|c||}{$(10^9-10^{13})~{\rm GeV}$} & EP& $1.1$
\\ \hline
\ref{om}-${\sf (c_2)}$ & \multicolumn{2}{|c||}{$10^{-10}-10^{-6}$}
& EP& $0.85$\\
&\multicolumn{2}{|c||}{$10^{-6}-1$} & EP& $1.1$
\\ \hline
\end{tabular}
\end{center}\vspace*{-.155in}
\caption{\sl The type of $\tilde\chi$-production and the chosen
$\delta_{\rm RH}$'s for various ranges of the $x$-axis para-meters
and $m_{\tilde\chi}$'s in Fig. \ref{om}.}
\end{table}

The validity of our semi-analytical approach can be tested by
comparing its results for $\Omega_{\tilde\chi}h^2$ with those
obtained by the numerical solution of Eqs (\ref{nf})-(\ref{nx}).
In addition, useful conclusions can be inferred for the behavior
of $\Omega_{\tilde\chi}h^2$ as a function of our free parameters
and the regions where each ${\tilde\chi}$-production mechanism can
be activated.

Our results are presented in Fig. \ref{om}. The solid lines are
drawn from our numerical code, whereas crosses are obtained by
employing (with $\delta_{\rm F}=1$) Eq. (\ref{nonBEsol}
[\ref{NBEsol}]) for [non] EP and solving numerically Eq.
(\ref{BEfRD})  with a convenient $\delta_{\rm RH}$ (comments on
the validity of Eq. (\ref{BEfan}) are given, too). The type of
${\tilde\chi}$-production and the chosen $\delta_{\rm RH}$'s for
the parameters used in Fig. \ref{om} are arranged in Table 2.

The light [normal] grey lines and crosses correspond to
$m_{\tilde\chi}=200~[500]~{\rm GeV}$. In Fig. \ref{om}-${\sf (a_1,
b_1, c_1~[a_2, b_2, c_2])}$, we fixed $\langle\sigma
v\rangle=10^{-12}~[10^{-8}]~{\rm GeV^{-2}}$. We design
$\Omega_{\tilde\chi}h^2$ versus:

$\bullet$  $T_{\rm RH}$, in Fig. \ref{om}-${\sf (a_1\ [a_2])}$ for
$N_{\tilde\chi}=10^{-6}~[10^{-3}]$ and $m_\phi=10^6~{\rm GeV}$.
Taking into account Table 2, also, we deduce that non-EP is
accommodated for lower $\langle\sigma v\rangle$ and $T_{\rm RH}$
and higher $m_{\tilde\chi}$, since in these cases, $Y_n~[Y_N]$ in
Eq. (\ref{BEsolneq1} [\ref{BEsolneq2}]) decreases, facilitating
the satisfaction of Eq. (\ref{cond}) for non-EP. The consideration
of $N_{\tilde\chi}$ allows us to produce acceptable results for
$\Omega_{\tilde\chi}h^2$ even for non-EP, in contrast with Ref.
\cite{fornengo}, where $N_{\tilde\chi}=0$. When
$\Omega_{\tilde\chi}h^2$ increases with $T_{\rm RH}$, Eq.
(\ref{BEfan}{\sf b}) works well. However, for larger
$\langle\sigma v\rangle$ and/or $T_{\rm RH}$,
$\Omega_{\tilde\chi}h^2$ decreases, when $T_{\rm RH}$ increases
(Fig. \ref{om}-${\sf (a_2)}$). None of Eqs (\ref{BEfan}) works in
this regime and so, the numerical solution of Eq. (\ref{BEfRD}) is
indispensable. For $m_{\tilde\chi}=200~[500]~{\rm GeV},
\langle\sigma v\rangle=10^{-8}~{\rm GeV^{-2}}$ and $T_{\rm
RH}>8.8~[22.5]~{\rm GeV}$, ${\tilde\chi}$ de-couples for
$x<x_{_{\rm RH}}$ (see sec. \ref{omega}).

$\bullet$ $m_\phi$, in Fig. \ref{om}-${\sf(b_1~[b_2])}$ for
$N_{\tilde\chi}=10^{-6}~[10^{-3}]$ and $T_{\rm RH}=0.05~[5]~{\rm
GeV}$. In both cases, $\Omega_{\tilde\chi}h^2$ decreases as
$m_\phi$ increases. Eq. (\ref{BEfan}${\sf b~[a]}$) gives reliable
results for $m_\phi>10^{9}$ and $m_{\tilde\chi}=500~[200]~{\rm
GeV}$ in Fig. \ref{om}-${\sf(b_2)}$ and for the parameters of Fig.
\ref{om}-${\sf(b_1)}$.

$\bullet$ $N_{\tilde\chi}$, in Fig. \ref{om}-${\sf (c_1~[c_2])}$
for $T_{\rm RH}=0.05~[5]~{\rm GeV}$ and $m_\phi=10^6~{\rm GeV}$.
We observe that $\Omega_{\tilde\chi}h^2$ increases with
$N_{\tilde\chi}$. Eq. (\ref{BEfan}${\sf b~[a]}$) gives reliable
results for $N_{\tilde\chi}<10^{-6}$ and
$m_{\tilde\chi}=500~[200]~{\rm GeV}$ in Fig. \ref{om}-${\sf(c_2)}$
and for the parameters of Fig. \ref{om}-${\sf(c_1)}$.
\addtolength{\textheight}{1.cm}
\newpage

%%%%%%%%%%%%%%%%%%%%%%%%%%%%%%%%%%%%%%%%%%%%%%%%%%%%%%%%%%%%%%%%%%%%
\begin{figure}[!h]\vspace*{-.15in}
\hspace*{-.71in}
\begin{minipage}{8in}
\epsfig{file=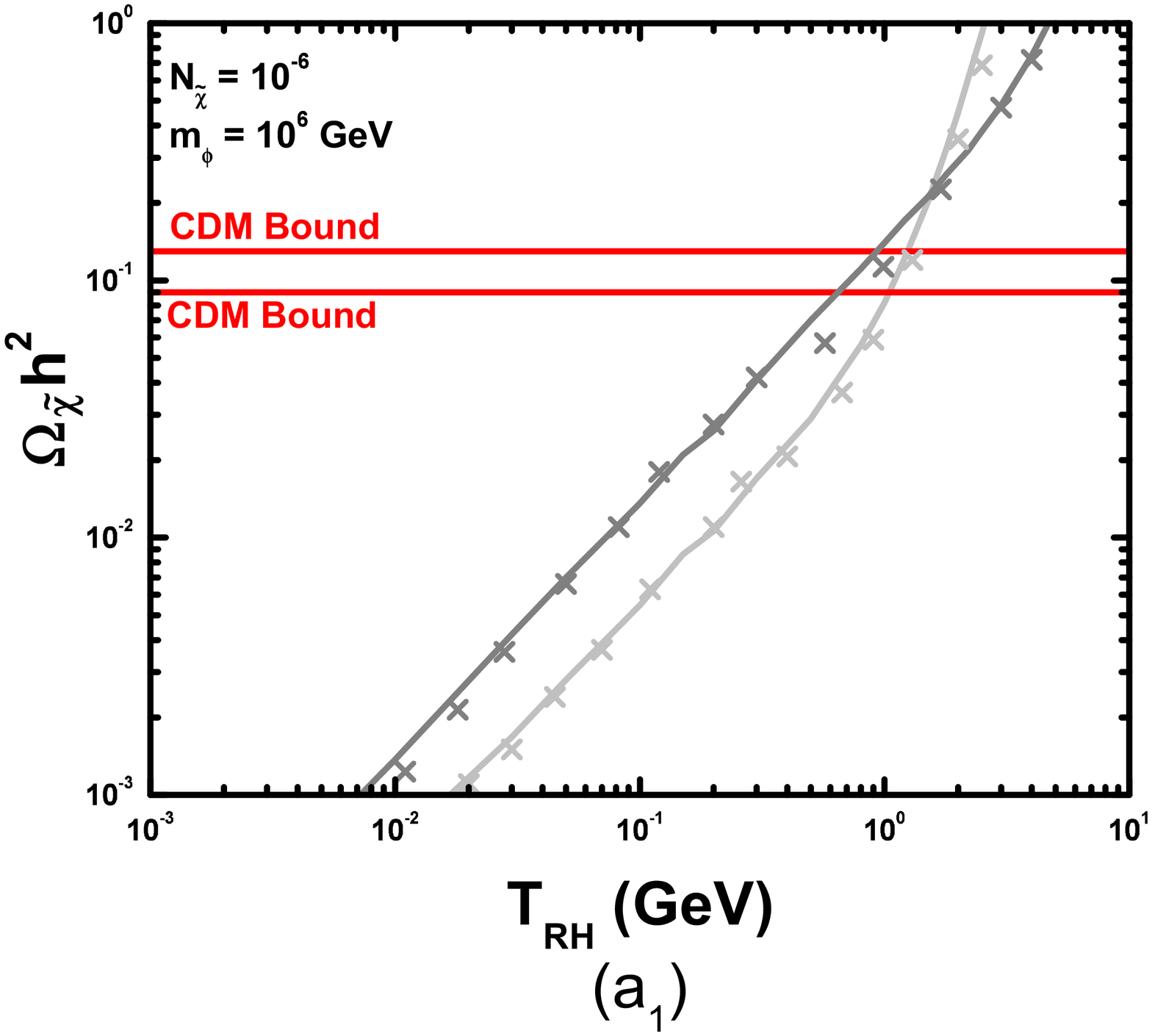,height=3.8in,angle=-90} \hspace*{-1.37 cm}
\epsfig{file=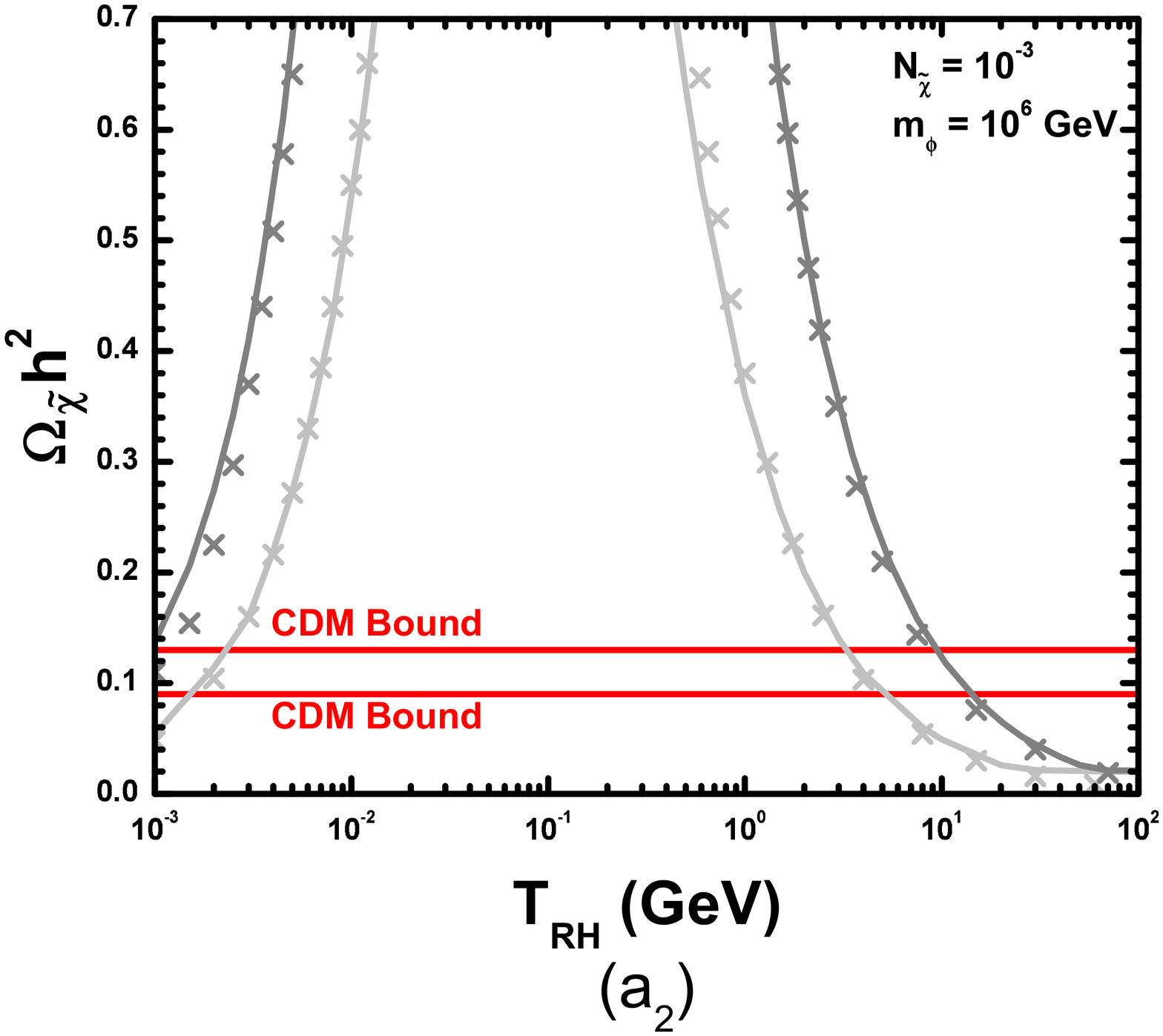,height=3.8in,angle=-90} \hfill
\end{minipage}\vspace*{-.01in}
\hfill\hspace*{-.71in}
\begin{minipage}{8in}
\epsfig{file=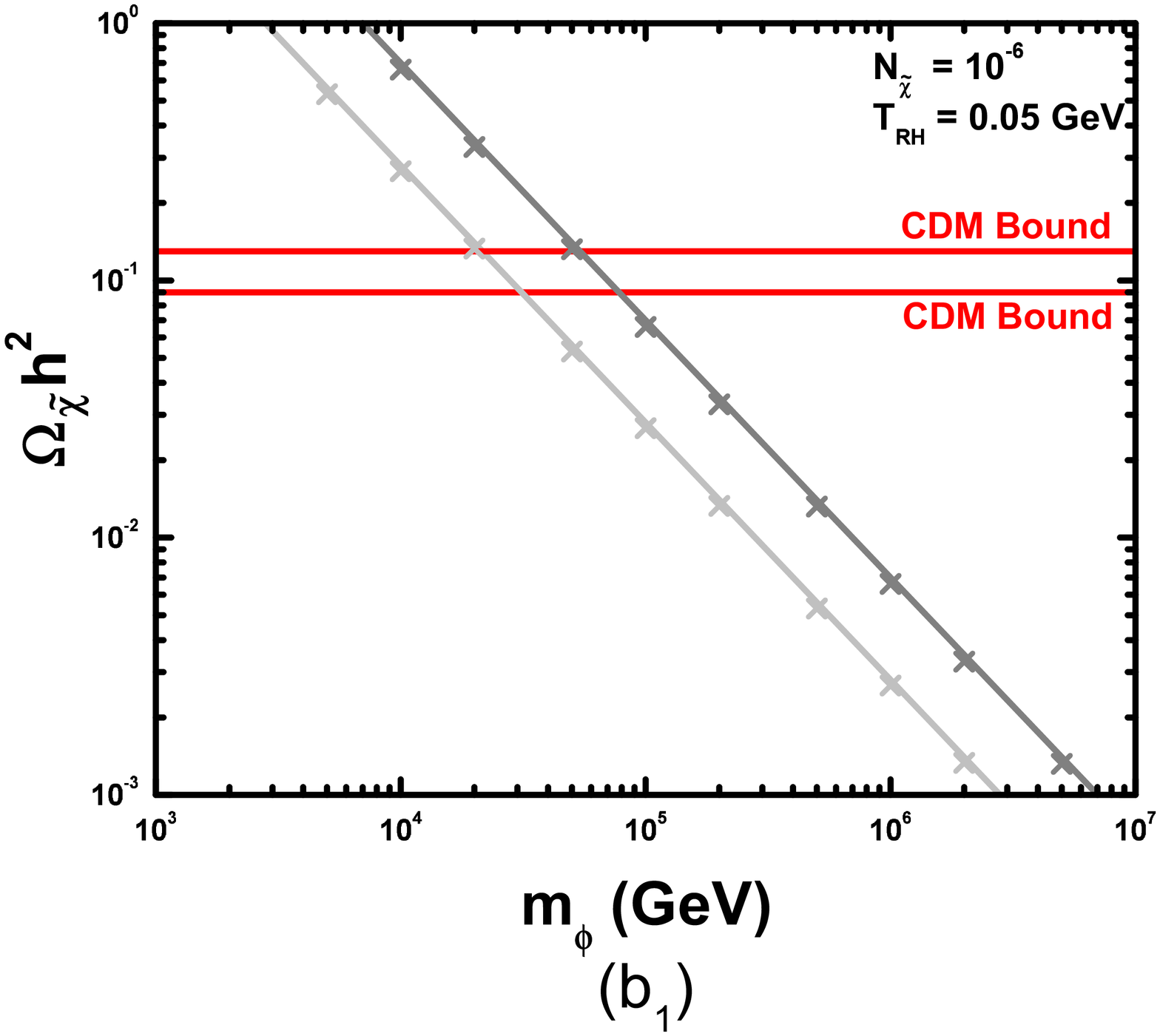,height=3.8in,angle=-90} \hspace*{-1.37 cm}
\epsfig{file=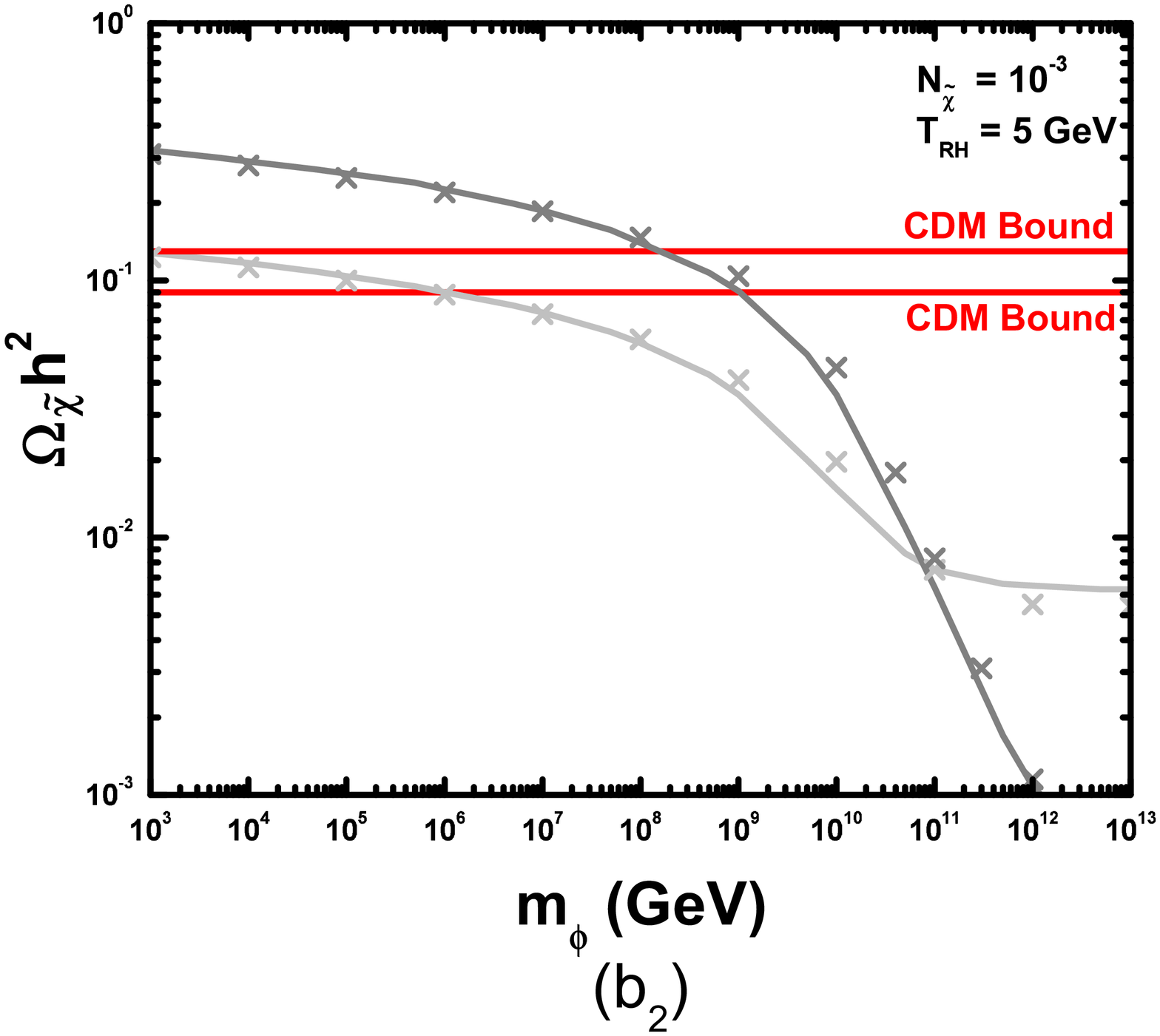,height=3.8in,angle=-90} \hfill
\end{minipage}\vspace*{-.01in}
\hfill\hspace*{-.71in}
\begin{minipage}{8in}
\epsfig{file=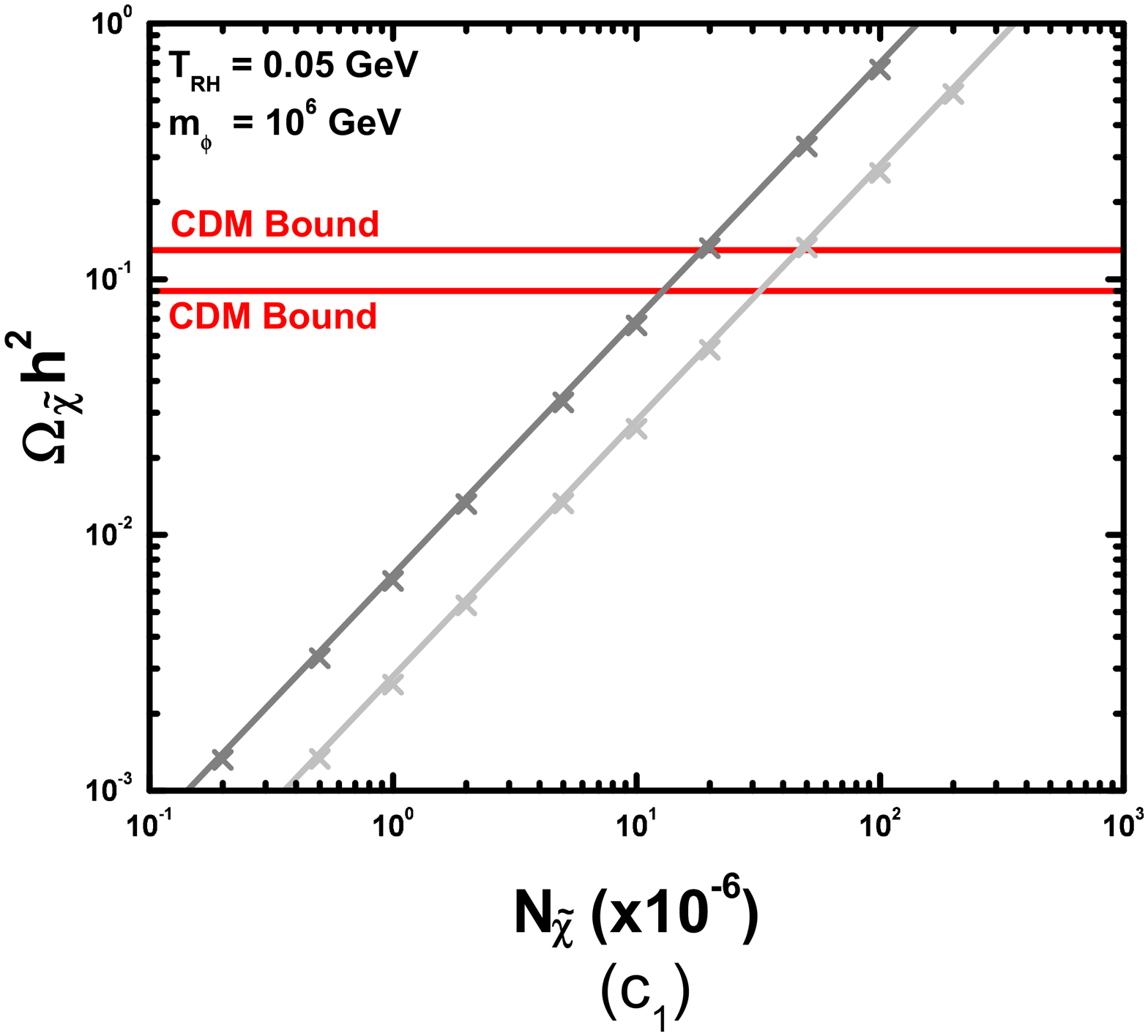,height=3.8in,angle=-90} \hspace*{-1.37 cm}
\epsfig{file=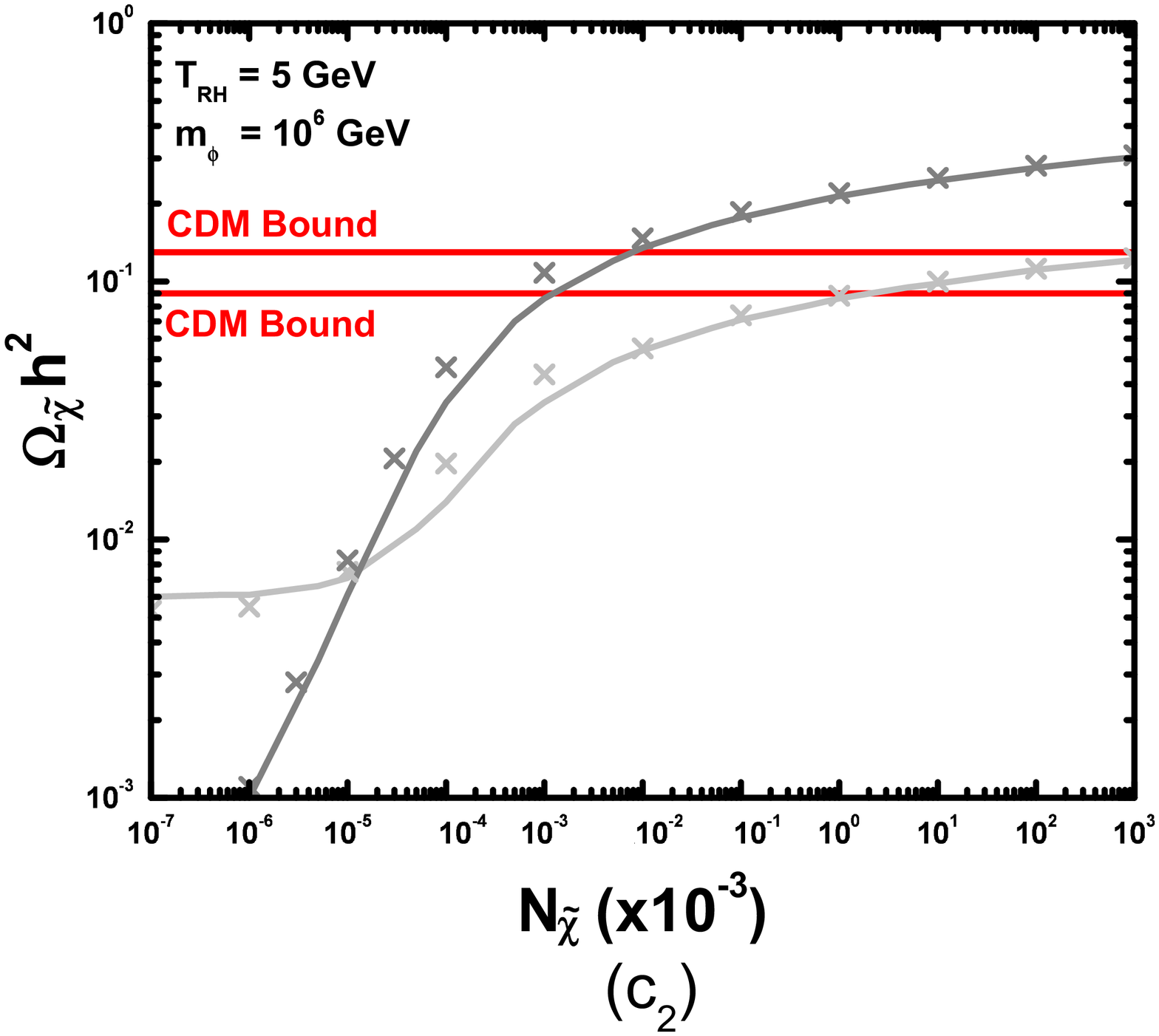,height=3.8in,angle=-90} \hfill
\end{minipage}\vspace*{-.05in}
\hfill \caption[]{\sl $\Omega_{\tilde\chi}h^2$ versus  $T_{\rm RH}
~{\sf ( a_1, a_2)}$, $m_\phi~{\sf(b_1, b_2)}$ and
$N_{\tilde\chi}~{\sf (c_1, c_2)}$ for fixed (indicated in the
graphs) $N_{\tilde\chi}$ and $m_\phi$, $N_{\tilde\chi}$ and
$T_{\rm RH}$, $T_{\rm RH}$ and $m_\phi$ correspondingly. We take
$m_{\tilde\chi}=200~[500]~{\rm GeV}$ (light [normal] grey lines
and crosses) and $\langle\sigma v\rangle=10^{-12}~[10^{-8}]~{\rm
GeV^{-2}}~{\sf (a_1, b_1, c_1~[a_2, b_2, c_2])}$. The solid lines
[crosses] are obtained by our numerical code [semi-analytical
expressions].}\label{om}
\end{figure}
%%%%%%%%%%%%%%%%%%%%%%%%%%%%%%%%%%%%%%%%%
\addtolength{\textheight}{-1.cm}
\newpage

Comparing Fig. \ref{om}-${\sf(b_1~[b_2])}$ and Fig. \ref{om}-${\sf
(c_1~[c_2])}$, it can be induced that $\Omega_{\tilde\chi}h^2$
remains invariant for constant $N_{\tilde\chi}m^{-1}_\phi$. This
can be understood by the observation that $y_{_N}$ in Eq.
(\ref{YN}) is proportional to this quantity. The adjustment of
$\delta_{\rm RH}$ turns out to be crucial, in order for the
results from the semi-analytical treatment to approach the
numerical ones. However, this kind of uncertainties is unavoidable
to such approximations (see Ref. \cite{riotto}). Note that the
accuracy of our approximations increases as
$N_{\tilde\chi}m^{-1}_\phi$ is reduced.

\subsection{A{\ssz LLOWED} R{\ssz EGIONS}}
\label{NTR}

\hspace{.562cm} Requiring $\Omega_{\tilde\chi}h^2$ to be confined
in the cosmologically allowed range of Eq. (\ref{cdmb}), one can
restrict the free parameters. The data is derived exclusively by
the numerical program.

Our results are presented in Fig. \ref{regions}. The light
[normal] grey regions are constructed for
$m_{\tilde\chi}=200~[500]~{\rm GeV}$. In Fig. \ref{regions}-${\sf
(a_1, b_1, c_1~[a_2, b_2, c_2])}$, we fixed $\langle\sigma
v\rangle=10^{-12}~[10^{-8}]~{\rm GeV^{-2}}$. We display the
allowed regions on the:

$\bullet$ $N_{\tilde\chi}-T_{\rm RH}$ plane, in Fig.
\ref{regions}-${\sf (a_1, a_2)}$ for $m_\phi=10^6~{\rm GeV}$.
Since $\Omega_{\tilde\chi}h^2$ increases with $T_{\rm RH}$ for low
$\langle\sigma v\rangle$ 's, the increase of $T_{\rm RH}$ requires
diminution of $N_{\tilde\chi}$ in Fig. \ref{regions}-${\sf
(a_1)}$. On the contrary, in Fig. \ref{regions}-${\sf (a_2)}$,
when decrease of $\Omega_{\tilde\chi}h^2$ is achieved with
increase of $T_{\rm RH}$ for  $T_{\rm RH}\gtrsim 2~{\rm GeV}$,
augmentation of $N_{\tilde\chi}$ is needed for increasing $T_{\rm
RH}$.

$\bullet$ $N_{\tilde\chi}-m_\phi$ plane, in Fig.
\ref{regions}-${\sf(b_1~[b_2])}$ for $T_{\rm RH}=0.05~[5]~{\rm
GeV}$. Since in both cases, $\Omega_{\tilde\chi}h^2$
in[de]-creases as $N_{\tilde\chi}~ [m_\phi]$ increases, increase
of $N_{\tilde\chi}$ entails increase of $m_\phi$.

$\bullet$ $T_{\rm RH}-m_\phi$ plane, in Fig. \ref{regions}-${\sf
(c_1~[c_2])}$ for $N_{\tilde\chi}=10^{-6}~[10^{-3}]$. Since for
$T_{\rm RH}\lesssim 2~{\rm GeV}$, $\Omega_{\tilde\chi}h^2$
increases with $T_{\rm RH}$ (see Fig. \ref{om}-${\sf (a_1, a_2)}$)
$m_\phi$ is to increase with $T_{\rm RH}$, as in Fig.
\ref{regions}-${\sf (c_1)}$. Similar region exists, also, for the
parameters used in Fig. \ref{regions}-${\sf (c_2)}$. However, for
the sake of illustration, in Fig. \ref{regions}-${\sf (c_2)}$ we
concentrated on the regions with $T_{\rm RH}\gtrsim 2~{\rm GeV}$.
In these, increase of $m_\phi$ is dictated with decrease of
$T_{\rm RH}$, since the latter causes increase of
$\Omega_{\tilde\chi}h^2$, as shown in Fig. \ref{om}-${\sf (a_2)}$.
The upper [lower] bounds of the allowed areas are derived from the
saturation of Eq. (\ref{cdmb}${\sf a~[b]}$) (inversely to all
other cases).

Note, finally, that in the standard paradigm,
$\Omega_{\tilde\chi}h^2$ is $(m_\phi, N_{\tilde\chi},T_{\rm
RH})$-independent and turns out to be almost
$143-146~[0.022-0.023]$ for $\langle\sigma
v\rangle=10^{-12}~[10^{-8}]~{\rm GeV^{-2}}$ and $m_{\rm LSP}$
varying within the range of Eq. (\ref{parb}{\sf a}). This is
easily extracted by solving Eq. (\ref{nx}) with $N_{\tilde\chi}=0$
and only a RD background in Eq. (\ref{Hini}) with $g_{\rho\ast}$
fixed at $x_{_{\rm F}}m_{\tilde\chi}$ with $17\lesssim
x^{-1}_{_{\rm F}}\lesssim26$. We checked that this result agrees
with this obtained by solving Eq. (76) of Ref. \cite{edjo}.

\section{C{\ftn ONCLUSIONS}-O{\ftn PEN} I{\ftn SSUES}}\label{con}

\hspace{.562cm} We considered a deviation from the standard
cosmological situation according to which the CDM candidate,
$\tilde\chi$ decouples from the plasma (i) during the RD era (i.e.
after reheating) (ii) being in equilibrium (iii) produced through
thermal scatterings. On the contrary, we assumed that $\tilde\chi$
decoupling occurs (i$^\prime$) during a decaying-massive-particle,
$\phi$, dominated era (and mainly before reheating) (ii$^\prime$)
being or not in chemical equilibrium with the thermal bath
(iii$^\prime$) produced by thermal scatterings and directly from
the $\phi$ decay.

We solved the problem (i) numerically, integrating the relevant
system of the differential equations (ii) semi-analytically,
producing approximate relations for the cosmological evolution
before reheating and solving the properly re-formulated Boltzmann
equations.

\addtolength{\textheight}{1.cm}
\newpage

%%%%%%%%%%%%%%%%%%%%%%%%%%%%%%%%%%%%%%%%%%%%%%%%%%%%%%%%%%%%%%%%%%%%
\begin{figure}[!ht]\vspace*{-.15in}
\hspace*{-.71in}
\begin{minipage}{8in}
\epsfig{file=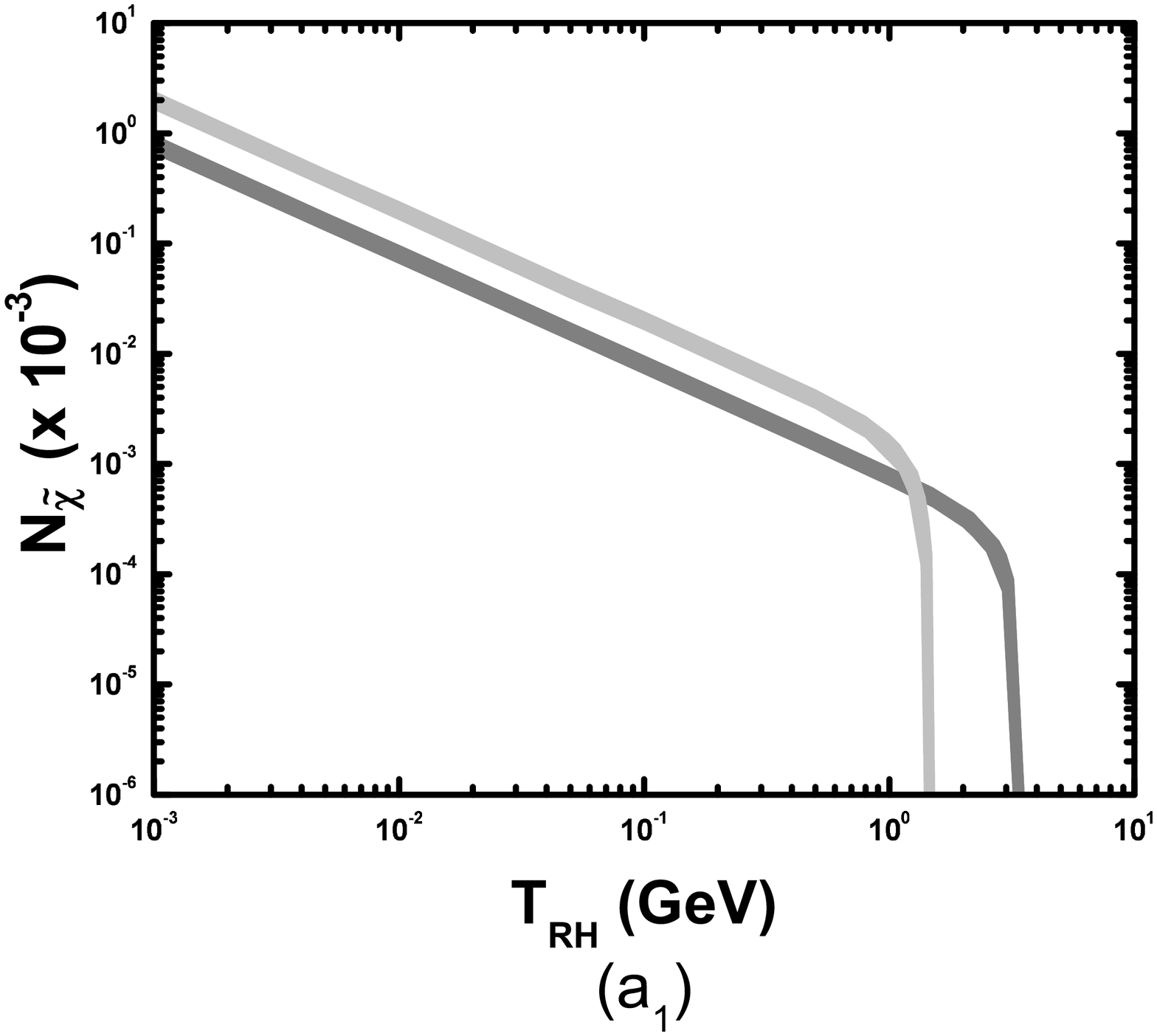,height=3.8in,angle=-90} \hspace*{-1.37 cm}
\epsfig{file=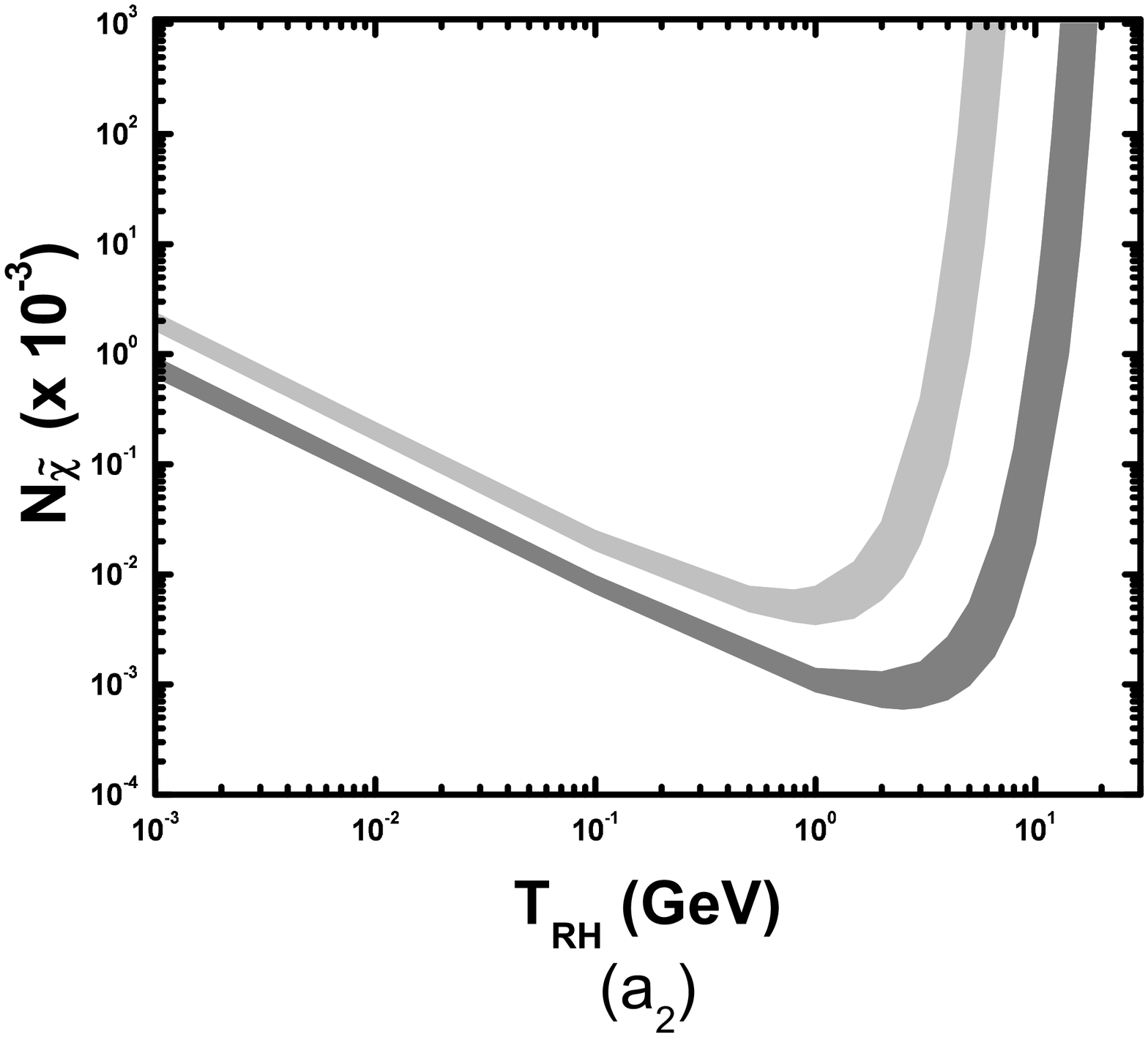,height=3.8in,angle=-90} \hfill
\end{minipage}\vspace*{-.01in}
\hfill\hspace*{-.71in}
\begin{minipage}{8in}
\epsfig{file=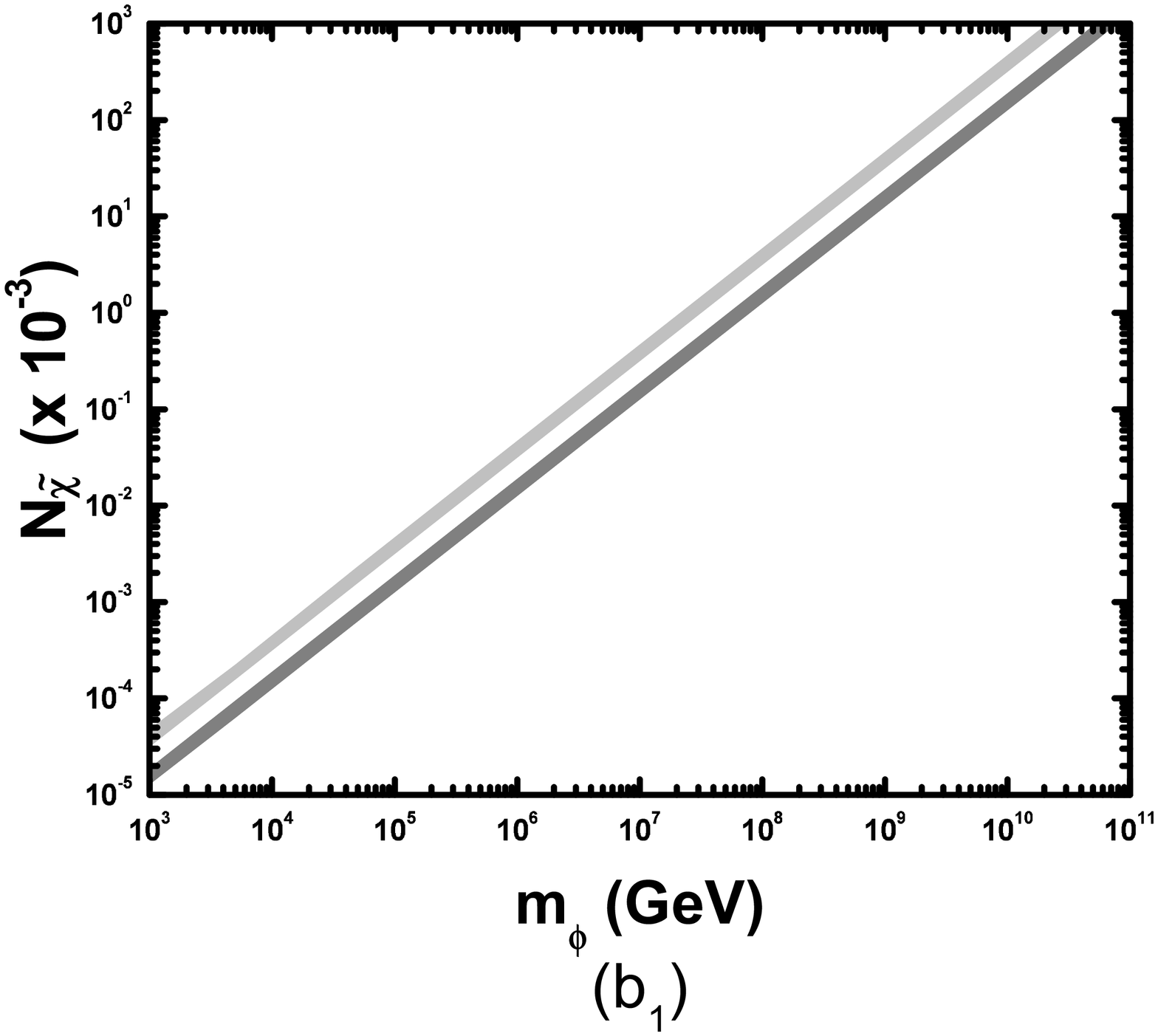,height=3.8in,angle=-90} \hspace*{-1.37 cm}
\epsfig{file=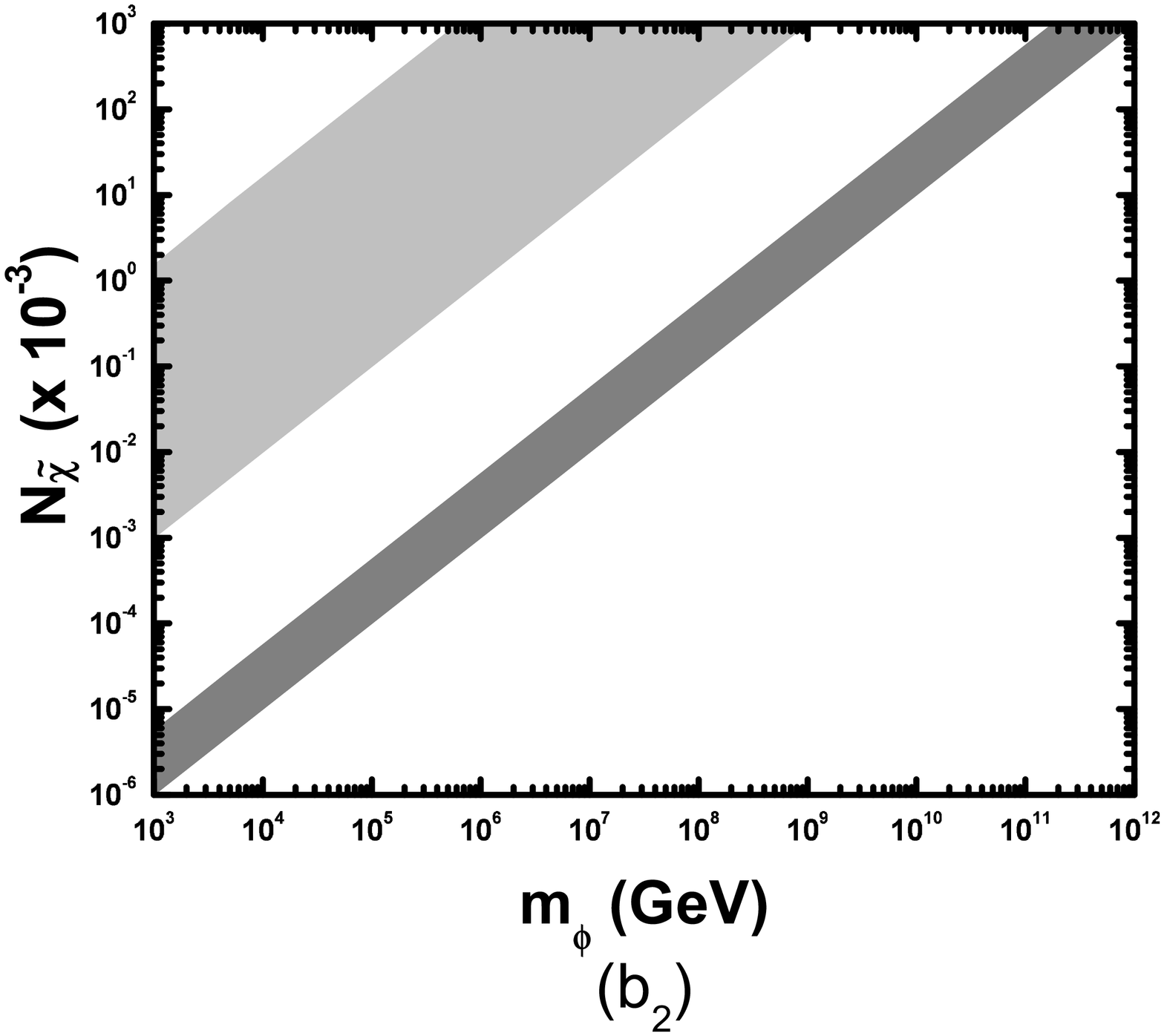,height=3.8in,angle=-90} \hfill
\end{minipage}\vspace*{-.01in}
\hfill\hspace*{-.71in}
\begin{minipage}{8in}
\epsfig{file=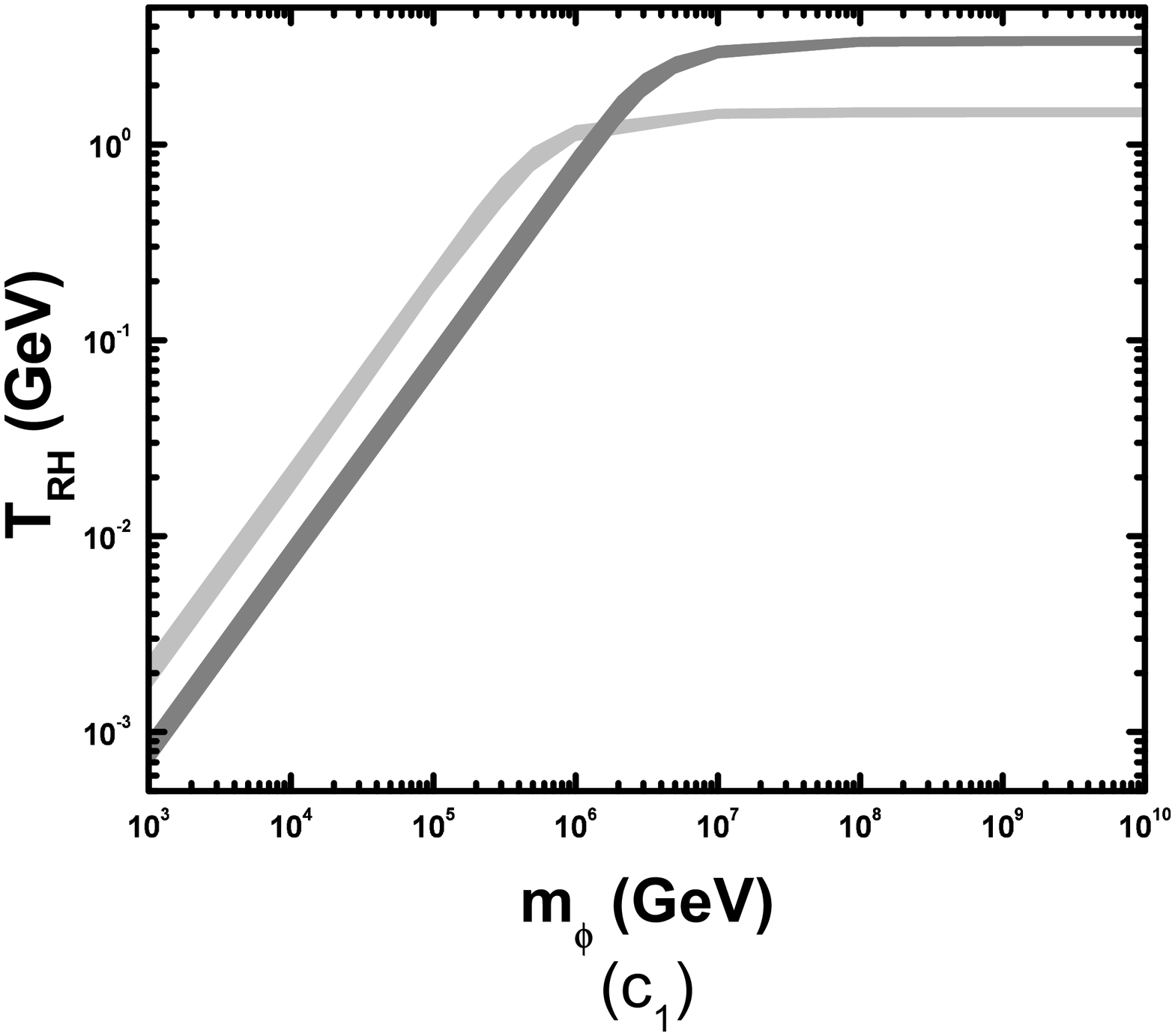,height=3.8in,angle=-90} \hspace*{-1.37 cm}
\epsfig{file=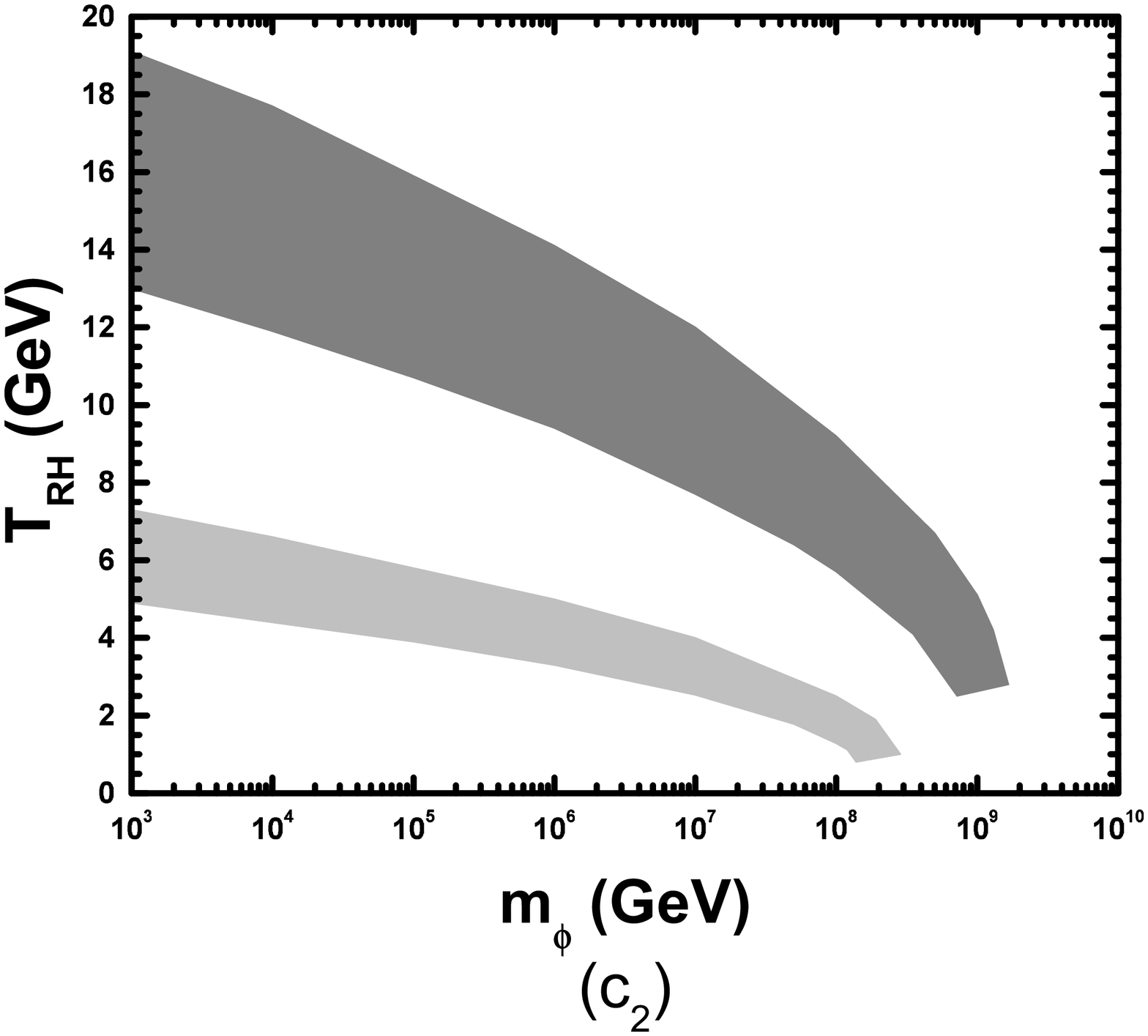,height=3.8in,angle=-90} \hfill
\end{minipage}\vspace*{-.05in}
\hfill \caption[]{\sl Regions allowed by Eq. (\ref{cdmb}) on the
$T_{\rm RH}-N_{\tilde\chi}$ plane ${\sf (a_1, a_2)}$ for
$m_\phi=10^6~{\rm GeV}$, $m_\phi-N_{\tilde\chi}$ plane ${\sf ( b_1
~[b_2])}$ for $T_{\rm RH}=0.05~[5]~{\rm GeV}$ and $m_\phi-T_{\rm
RH}$ plane ${\sf (c_1~[c_2])}$ for $N_{\tilde\chi}=10^{-6[3]}$. We
take $m_{\tilde\chi}=200~[500]~{\rm GeV}$ (light grey [normal
grey] areas) and $\langle\sigma v\rangle=10^{-12}~[10^{-8}]~{\rm
GeV^{-2}}~{\sf (a_1, b_1, c_1~[a_2, b_2, c_2])}$.} \label{regions}
\end{figure}
%%%%%%%%%%%%%%%%

\addtolength{\textheight}{-1.cm}
\newpage

The second way facilitates the understanding of the problem and
gives, in most cases, sufficiently accurate results, provided a
suitable $\delta_{\rm RH}$ is chosen. The variation of
$\Omega_{\tilde\chi}h^2$ w.r.t our free parameters $(m_\phi,
N_{\tilde\chi},T_{\rm RH})$ was investigated and regions
consistent with the present CDM bounds are constructed, using
$m_{\tilde\chi}$'s and $\langle\sigma v\rangle$'s commonly allowed
in SUSY models.

These scenaria obviously let intact the SUSY parameter space but
require rather low $T_{\rm RH}$. This can be accommodated in AMSBM
\cite{moroi}, in models with intermediate scale unification
\cite{shaaban} and within the context of $q$-balls decay
\cite{fujii}. Also, low $T_{\rm RH}$ naturally arises in models of
thermal inflation  \cite{lyth} which, as a bonus, may overcome the
problem of unwanted relics (e.g., gravitino, moduli). Finally,
restrictions on $T_{\rm RH}$ arising from  baryogenesis and
neutrino cosmology have been, also, studied in Refs \cite{riotto1,
neutrino}.

Our formalism can be easily extended to include coannihilations
and pole effects. Therefore, it can become applicable for the
calculation of $\Omega_{\tilde\chi}h^2$ in the context of specific
SUSY models. Also, these scenaria can assist us to the reduction
of $\Omega_{\tilde\chi}h^2$ in cases, where it turns out to be
even more enhanced than in the standard scenario, as in the
presence of Quintessence \cite{salati}. Similar analysis of the
$\tilde\chi$-decoupling during the extra dimensional cosmological
evolution \cite{extra} may be also, possible.

\acknowledgments \hspace{.562cm} The author gratefully
acknowledges A. Riotto, S. Khalil  and A. Masiero for inspiring
discussions, G. Lazarides, T. Moroi and N.D. Vlachos for useful
correspondence, M. Drees, X. Zhang and S. Profumo for interesting
comments. This work was supported by European Union under the RTN
contract HPRN-CT-2000-00152.

\end{document}